\documentclass[12pt]{article}

\usepackage{amsmath,amssymb,epsfig,amsfonts}
\usepackage{graphicx}
\usepackage{cite}
\usepackage[usenames,dvipsnames]{xcolor}
\usepackage{setspace}
\usepackage{lscape}
\usepackage{bm}
\usepackage{picture}
\usepackage{tikz}
\usepackage{hyperref}
\usepackage[top=.85in, left=0.6in, right=0.6in, bottom=1in]{geometry}
\usepackage[width=0.87\textwidth]{caption}

\newcommand{\ov}{\overline}

\makeatletter

\def\rr{\mathsf{q}}
\def\P{\mathbb{P}}

\def\zb{\bar{z}}
\def\mfH{{\mathfrak{H}}}
\def\Z{\mathbb{Z}}
\def\C{\mathbb{C}}
\def\R{\mathbb{R}}

\def\ie{{\it i.e.}}
\def\rank{\mathrm{rank}}
\def\lg{\mathfrak{g}}
\def\cT{\mathcal{T}}
\def\lt{\mathfrak{t}}

\def\cI{\mathcal{I}}
\def\xib{\bar{\xi}}

\def\mm{\mathsf{m}}
\def\mfI{{\mathfrak{I}}}
\def\cW{\mathcal{W}}
\def\cV{\mathcal{V}}

\newcommand{\bZ}{\mathbb{Z}}

\newcommand{\bP}{\mathbb{P}}
\newcommand{\bR}{\mathbb{R}}

\newcommand{\cN}{\mathcal{N}}
\newcommand{\cM}{\mathcal{M}}
\newcommand{\cZ}{\mathcal{Z}}

\newcommand{\cF}{\mathcal{F}}

\newcommand{\cO}{\mathcal{O}}

\newcommand{\cL}{\mathcal{L}}
\newcommand{\cC}{\mathcal{C}}

\newcommand{\ep}{\epsilon}

\def\Res{\mathop{\mathrm{Res}}\limits}

\def\tr{\mathop{\mathrm{tr}}\nolimits}

\def\ov#1{{\overline{#1}}}

\def\unit{{1\kern-.65ex {\rm l}}}
\def\1{{1\kern-.65ex {\rm l}}}

\def\bt{{\widetilde{b}}}

\def\xb{{\overline{x}}}
\def\zb{{\overline{z}}}

\renewcommand{\bt}{\bm{\tau}}
\newcommand{\bQ}{\bm{Q}}
\newcommand{\bth}{\bm{\theta}}
\newcommand{\br}{\bm{r}}

\newcommand{\Zc}{Z_{\text{class}}}
\newcommand{\Zg}{Z_{\text{gauge}}}

\newcount\hour \newcount\minute
\hour=\time \divide \hour by 60
\minute=\time
\def\now{%
\ifnum \hour<13
  \ifnum \hour=0 \advance \hour by 12 \number\hour:\else \number\hour:\fi%
     \ifnum \minute<10 0\fi%
     \number\minute%
\ A.M.%
\else \advance \hour by -12 \number\hour:%
  \ifnum \minute<10 0\fi%
  \number\minute%
  \ P.M.%
\fi%
}

\makeatother

\begin{document}

\baselineskip=18pt  
\numberwithin{equation}{section}  
\allowdisplaybreaks  

\thispagestyle{empty}

\vspace*{-2cm} 
\begin{flushright}
{\tt UCSB Math 2013-05} \\
{\tt NSF-KITP-13-076}
\end{flushright}

\vspace*{0.5cm} 
\begin{center}
 {\LARGE  
New Methods for Characterizing Phases of \\
2D Supersymmetric Gauge Theories}
 \vspace*{1.0cm}

James Halverson$^1$, Vijay Kumar$^1$, and David R. Morrison$^{2,3}$

\vspace{1.5cm}

$^1$Kavli Institute for Theoretical Physics, University of California\\
Santa Barbara, CA 93106, USA \\ \vspace{.3cm}  
$^2$Department of Mathematics, University of California\\ 
Santa Barbara, CA 93106, USA \\ \vspace{.3cm} 
$^3$Department of Physics, University of California\\
Santa Barbara, CA 93106, USA

\vspace*{0.5cm}
\noindent {\footnotesize \texttt{jim@kitp.ucsb.edu, vijayk@kitp.ucsb.edu, drm@math.ucsb.edu}}

\end{center}
\vspace*{1cm}

\abstract{

\vspace{.5cm} 

We study the physics of two-dimensional $\cN=(2,2)$
gauged linear sigma models (GLSMs) via the two-sphere partition
function.
We show that the classical phase boundaries separating distinct GLSM phases, which are described by the secondary fan construction for abelian GLSMs, are completely encoded in the analytic structure of the partition function. 
The partition function of a non-abelian GLSM can be obtained as a limit from an abelian theory; we utilize this fact to show that the phases of non-abelian  GLSMs can be obtained from the secondary fan of the associated abelian GLSM.
We prove that the partition function of any abelian GLSM satisfies a set of linear differential equations; these reduce to the familiar $A$-hypergeometric
system of Gel'fand, Kapranov, and Zelevinski  for GLSMs describing complete intersections in toric varieties.
We develop a set of conditions that are necessary 
for a GLSM phase to admit an interpretation as the low-energy limit of a
 non-linear sigma model with a Calabi--Yau threefold target space.
Through the application of these criteria we discover a class of GLSMs with novel geometric phases corresponding to Calabi--Yau manifolds that are branched double-covers of Fano threefolds. These criteria provide a promising approach for constructing new Calabi--Yau geometries.
}

\vspace{0.5in}
\noindent
\today

\clearpage

\parskip 0.1in

\setstretch{1}

\section{Introduction}

The gauged linear sigma model (GLSM) has provided many insights on the
space of Calabi--Yau threefolds \cite{Witten:1993yc}. It is a two-dimensional
supersymmetric gauge theory with charged matter fields which, if the
degrees are constrained correctly, flows to a
supersymmetric conformal field theory (SCFT) at low energies. 
For a large class of Calabi--Yau threefolds $X$, 
with a judicious choice of gauge group, matter content and interactions
one can design a GLSM that flows to the same SCFT as a non-linear sigma
model (NLSM) whose target space is $X$.
 Various geometric properties of $X$, such as its quantum moduli space (the space of marginal
deformations of the SCFT), can be analyzed in terms of GLSM physics. For
example, the properties of distinct classical phases in the GLSM determine the
nature of singular points in the (compactified) moduli space \cite{Witten:1993yc}.

Recently, it was conjectured that the two-sphere partition function of
a GLSM --- which was computed independently in
\cite{Benini:2012ui,Doroud:2012xw} --- allows one to determine the
Gromov-Witten invariants of Calabi--Yau
threefolds\cite{Jockers:2012dk}. A compelling physical argument for
this conjecture has been given in \cite{Gomis:2012wy}. For Calabi--Yau
threefolds\footnote{There has also been an extension to Calabi--Yau fourfolds \cite{arXiv:1302.3760}.} that can be modeled with a GLSM,
this new approach allows one to compute these invariants and the exact
metric on moduli space without the use of mirror symmetry
\cite{Jockers:2012dk,Park:2012nn,Sharpe:2012fk}.

Given these previous successes, in this paper we use the two-sphere
partition function as a lens for studying GLSM physics. We focus our
attention on three problems: defining the (classical) boundaries separating distinct GLSM
phases, determining whether a given GLSM phase is geometric, and
defining a system of differential equations on the moduli space of the
low-energy SCFT.

The partition function takes the form of a contour integral for both
abelian and non-abelian GLSMs. We
show that as far as the partition function is concerned, the physics
of the non-abelian theory can be extracted from an abelian theory that
we call the \emph{associated Cartan theory}. 
 This proposal resonates with an early proposal made by Hori and Vafa \cite{Hori:2000kt}, is related to some developments in the mathematical study of Gromov-Witten invariants \cite{Kim:06}, and also appeared more recently in the context of the partition function  \cite{Benini:2012ui}.
The analysis of the physics of non-abelian GLSMs can be
quite involved, as discussed in \cite{Hori:2006dk} for example. 
We utilize the associated Cartan theory to extend
many of our results for abelian theories to non-abelian theories.

We demonstrate that classical phase boundaries in the GLSM are encoded
in the analytic structure of the partition function integrand. We show
that the boundaries as determined by the two-sphere partition function
reproduce the known results (the structure of the secondary fan) for abelian GLSMs. Moreover, we show
that the phase boundaries for non-abelian GLSMs can be obtained from
the secondary fan of the associated Cartan theory.

We show that the partition function can be used as a tool to determine
whether a given phase in a GLSM is geometric, \ie, whether the
low-energy physics is approximated by an NLSM with a smooth Calabi--Yau
threefold as its target space. This question has traditionally been answered by a
study of the solutions to the D-term and F-term equations \cite{Witten:1993yc, Aspinwall:1994cf}, which can
be non-trivial as illustrated by the Hori--Tong analysis of the GLSM for
R\o{}dland's Pfaffian Calabi--Yau threefold \cite{Hori:2006dk}.  We
formulate a set of conditions, which we refer to as the
\emph{geometric phase criteria}, that are necessary for a given GLSM
phase to be geometric. These conditions provide a valuable cross check
when the D-term and F-term analysis is subtle, as was the case in
\cite{Hori:2006dk}.

The GLSM, together with the geometric phase criteria, can be a
valuable tool to explore the space of Calabi--Yau threefolds. Rather
than starting with a known Calabi--Yau threefold $X$ and then constructing a
GLSM that models $X$ in one of its phases, we start with a general
collection of GLSMs and focus our attention on the subcollection of GLSMs which satisfy the
geometric phase criteria. Apart from reproducing the list of known
Calabi--Yau threefolds, a closer analysis of this subcollection could
potentially lead to new constructions of Calabi--Yau threefolds.

We provide an instantiation of this approach by considering a collection of
GLSMs with gauge group $U(1)$. We find a twenty member subcollection that
satisfy the geometric phase criteria. Five of these examples
correspond to known Calabi--Yau threefolds that are complete
intersections in projective space. A closer analysis of the remaining
fifteen examples suggests that the phase in which the geometric
criteria are satisfied is a so-called ``hybrid phase.'' We then provide a physical
argument which shows that these fifteen examples are actually geometric
and correspond to double-covers of Fano threefolds branched over a
surface of appropriate degree. 
These examples are close relatives of a phase studied in  \cite{Caldararu:2007tc, Hellerman:2006zs} that occurs in a GLSM for a complete intersection of
four quadrics in $\P^7$. This GLSM phase was argued in those papers
to correspond to a
Calabi--Yau threefold geometry (a non-commutative resolution of a
double-cover of $\P^3$ branched over a singular octic surface).\footnote{See also the recent work that utilizes the two-sphere partition function to argue for such a geometric interpretation \cite{Sharpe:2012ji}.}  The
examples that we uncover are simpler, more traditional smooth
Calabi--Yau geometries that appear as hybrid phases in a GLSM.

The moduli space of an $N=2$ SCFT has a rich structure that is known
as special geometry \cite{deWit:1984pk, Periwal:1989mx, Strominger:1990pd, Candelas:1990pi, Craps:1997gp, MR717607, Freed:1997dp}. Such a moduli space of dimension $s$ carries an
$Sp(2s+2,\Z)$  bundle to which we can associate a set of
differential equations \cite{Cadavid:1991yh,Lerche:1991wm,Morrison:1991cd}
with regular singular points \cite{MR0417174}. 
The Picard--Fuchs differential equations that describe the variation of complex structure of Calabi--Yau threefolds are an example.
For threefolds that arise as complete intersections in
toric varieties, 
the solutions to the Picard--Fuchs equations are annihilated 
\cite{BatyrevDuke,BvS,Hosono:1993qy} by the differential operators of  the
$A$-hypergeometric system \cite{GKZ1, GKZ2} of Gel'fand, Kapranov, and
Zelevinski (GKZ). We show that one can associate a set of differential
equations, which we call the $A$-system, to \emph{any} GLSM with an
abelian gauge group. The $A$-system is a slight generalization of the
$A$-hypergeometric system and applies to a larger class of $N=2$
SCFTs.

The outline of the paper is as follows: section 2 discusses the
analytic structure of the two-sphere partition function and introduces
the associated Cartan theory; section 3 is a discussion of GLSM phases
from the partition function point of view; section 4 introduces the
geometric phase criteria, which are then applied in section 5 to a
collection of GLSMs to discover novel Calabi--Yau threefold geometries;
section 6 shows that we can associate a system of differential
equations to the moduli space of any abelian GLSM; we summarize our
findings and discuss possible future directions in section 7.

\section{Two-Sphere Partition Function}
\label{sec:two-sphere}
In this section we will review the two-sphere partition function of
$d=2$ $\cN=(2,2)$ GLSMs and discuss aspects of
its structure that will be useful in sections \ref{sec:phases} and
\ref{sec:geometries}. We also introduce the necessary notation that will be used in the rest of the paper.

We consider an $\cN=(2,2)$ GLSM with gauge group $G$, where $G$ is a
compact Lie group with associated Lie algebra $\lg$, and $n$ chiral
multiplets $\Phi_{i=1,\cdots, n}$ in irreducible representations $R_i$
of $G$. By an abuse of notation we will also use $R_i$ to denote the
 representation of the corresponding Lie algebra $\lg$.  We require the existence of
a $U(1)_V$ R-symmetry and denote the R-charge of $\Phi_i$ by
$\rr_i$. Let $\lt_\mu$, $\mu = 1, \cdots, \rank(\lg)$, denote the
generators of the Cartan subalgebra. We will assume that a non-empty
subset of these generators, $\{ \lt_1, \cdots, \lt_s\}$, commutes with
every element of the Lie algebra $\lg$. Without loss of generality we
have assumed these to be the first $s$ generators.  For each
$\lt_{a=1, \cdots, s}$ we have a Fayet-Iliopoulos (FI) parameter $r_a$
and a periodic theta angle $\theta_a$, which we write using vector notation as
$\br \in \bR^s$ and $\bth \in (\bR/(2\pi \Z))^s$. These parameters can
be naturally combined as $z_a = \exp(-2\pi r_a + i\theta_a) \in
\C^*$. Throughout the paper $s$ will denote the number of FI parameters
with the index $a=1, \cdots, s$, and the index $\mu = 1, \cdots,
\rank(\lg)$.

We will consider GLSMs that flow to superconformal field theories
(SCFTs) at low energies. A necessary condition for this is that the
axial $U(1)$ R-symmetry be non-anomalous \cite{Witten:1993yc}: 
$\sum_{i=1}^n \tr_{R_i}
(\mathfrak{t}) = 0$, for all $\lt \in \mathfrak{g}$. The central
charge of the resulting SCFT is given by  \cite{Silverstein:1994ih}:
$\hat{c} = c/3=\sum_{i=1}^n (1-\rr_i)\, dim(R_i) - \dim(\lg)$.

The two-sphere partition function in the Coulomb branch 
representation is given by \cite{Benini:2012ui,Doroud:2012xw}
\begin{equation}
  \label{eq:ZS2}
  Z_{S^2}(z_a, \zb_a) = \frac{1}{|\mathcal{W}|} \sum_{\bm{m} \in \Z^{\rank(\lg)}}\ \  \int \ \left( \  \prod_{\mu=1}^{\rank(\lg)} \frac{d \tau_\mu}{2 \pi i} \ \right)
  \,\,\, Z_{\text{class}} \,\,\, Z_{\text{gauge}} \prod_i \,\,Z_{\phi_i}\, ,
\end{equation}
where the classical factor $\Zc$ and the one-loop determinants $Z_{\phi_i}$ and $\Zg$ are given by 
\begin{align}
  \label{eq:Zc}
  \Zc &= e^{-4\pi \bm{r}\cdot \bt - i \bm{\theta} \cdot \bm{m}} = \prod_{a=1}^s \ (z_a)^{\tau_a-m_a/2} \ (\zb_a)^{\tau_a+m_a/2} \, , \\ 
  \Zg &=  \prod_{\alpha \in \mathcal{P}} \left( 
  \frac{(\alpha_\mu { m}_\mu)^2}{4} - 
  (\alpha_\mu  \tau_\mu)^2\right)\, ,\\
  Z_{\phi_i} &= \prod_{\rho^i \in R_i} \frac{\Gamma(\frac{\rr_i}{2}  - \rho^i_\mu ( \tau_\mu + \frac{ m_\mu}{2}))}{\Gamma(1-\frac{
\rr_i}{2} +\rho^i_\mu  ( \tau_\mu - \frac{ m_\mu}{2}))}\, .
\end{align}
In the above formulae $(\alpha_\mu) \in \mathcal{P} $, the set of positive roots of $\lg$, and the set of $(\rho^i_\mu)$ denotes the weights of the representation $R_i$ of $\lg$. Note that $\alpha_\mu$ and $\rho^i_\mu$ are vectors in the weight lattice $\cong \Z^{\mathrm{rank}(\lg)}$. $\mathcal{W}$ denotes the Weyl group of $\lg$ and $|\mathcal{W}|$, its cardinality. The boldface notation in \eqref{eq:Zc} is used throughout the paper as a shorthand for $s$-tuples. The integration is performed over the imaginary axis for each $\tau_\mu$. When the R-charges $\rr_i$ are all positive, which we will assume to be the case, there are no poles on the contour of integration. In some cases, we take a limit where the R-charges go to zero from the positive direction for simplicity.

Most of the discussion in this paper will focus on GLSMs where the gauge group $G$ is abelian. Note that $s = \rank(\lg) = \dim(\lg)$ in this case. We will denote the charges of the chiral fields by the vectors $\bm{Q}_i \in \Z^s$, $i = 1, \cdots, n$. The partition function in this case reads
\begin{equation}
 Z_{S^2}(z_a, \zb_a) = \sum_{\bm{m} \in \Z^s}\ \  \int \ \left( \  \prod_{a=1}^{s} \frac{d \tau_a}{2 \pi i} \  \ (z_a)^{\tau_a-m_a/2} \ (\zb_a)^{\tau_a+m_a/2} \right) 
 \prod_{i=1}^n
 \frac{\Gamma(\frac{\rr_i}{2}  - \bm{Q}_i\cdot ( \bm{\tau} + \frac{ \bm{m}}{2}))}{\Gamma(1-\frac{
\rr_i}{2} +\bm{Q}_i \cdot  ( \bm{\tau} - \frac{ \bm{m}}{2}))}\, .
\end{equation}

In a GLSM the complexified FI parameters, $z_a$, correspond to exact marginal parameters of the IR SCFT and parameterize a moduli space $\cM$. When the SCFT has a geometric interpretation --- as the low-energy limit of a non-linear sigma model (NLSM) with target space a Calabi--Yau threefold $X$--- the moduli space $\cM$ corresponds to (part of) the quantum K\"ahler moduli space of $X$.\footnote{In general, $\cM$ is a subset of the full K\"ahler moduli space, since extra exact marginal operators can be emergent at the IR fixed point.} It was conjectured in \cite{Jockers:2012dk} that the two-sphere partition function computes the exact K\"ahler potential, and hence the exact K\"ahler metric, on $\cM$. More precisely,
\begin{equation}
Z_{S^2} = \exp(-K(z_a, \zb_a))\, ,
\end{equation}
where $K$ is the K\"ahler potential on $\cM$. A compelling physical argument for this conjecture has been provided by Gomis and Lee \cite{Gomis:2012wy}.

\subsubsection*{Non-abelian GLSMs and their associated Cartan Theory}

It is apparent from the structure of \eqref{eq:ZS2} that the partition function of a non-abelian GLSM is closely related to the partition function of an \emph{abelian} GLSM. The maximal torus $\cT_G$ of $G$ is the gauge group of the abelian theory, with Lie algebra equal to the Cartan subalgebra $\mathfrak{u}(1)^{\oplus \rank(\lg)}$ of $G$. We will henceforth refer to this abelian theory as the \emph{associated Cartan theory}.
Each chiral multiplet $\Phi_i$
of the GLSM reduces to $\dim(R_i)$ chiral multiplets with charges
specified by the weights of the representation --- the vectors $(\rho^i) \in
\Z^{\rank(\lg)}$. Note that we can rewrite $\Zg$ as
\begin{equation}
\Zg = \nonumber \prod_{\alpha \in \mathcal{P}} \left( 
  \frac{(\alpha_\mu { m}_\mu)^2}{4} - 
  (\alpha_\mu  \tau_\mu)^2\right)\,  = \, \left(\prod_{\alpha \in \mathcal{P}} e^{i\pi \sum_{\mu} \alpha_\mu m_\mu}\right)\ \left(\prod_{\alpha \in \mathcal{R}} \frac{\Gamma(1 - \alpha_\mu ( \tau_\mu +\frac{ m_\mu}{2})}{\Gamma (\alpha_\mu ( \tau_\mu -\frac{ m_\mu}{2}))} \right)\, ,
\end{equation}
where $\mathcal{R}$ is the set of roots of $\lg$.  This leads us to include $\dim(\lg) - \rank(\lg)$ additional chiral multiplets with charges given by the roots 
$(\alpha_\mu)$ and R-charge $2$. The Cartan theory has a total of $\rank(\lg)$ FI parameters with an additional $\rank(\lg)-s$ parameters, denoted $(\tilde z_\mu)$, $\mu=s+1, \cdots, \rank(\lg)$, as compared to the original non-abelian theory. 
The partition functions of the two theories, by construction, are related as
\begin{equation}
Z_{S^2}^{NA} (z_a, \zb_a) = Z_{S^2}^{C}(z_a,\ \zb_a,\ \tilde{z}_\mu=e^{2\pi i \gamma_\mu},\ \bar{\tilde{z}}_\mu=e^{-2 \pi i \gamma_\mu})\, ,
\end{equation}
where $\gamma = \frac{1}{2} \sum_{\alpha \in \mathcal{P}} \alpha_\mu$
is the so-called Weyl vector. We use the labels $C$ and $NA$ to denote
the Cartan theory and the original non-abelian theory respectively.
We emphasize that the analytic structure of the $Z_{S^2}$ integrands
are identical in these two theories; the partition functions differ
only in the $\Zc$ factor according to the dependence on FI parameters.

The idea of associating an abelian GLSM to a non-abelian GLSM has been studied in the past starting with Hori and Vafa \cite{Hori:2000kt}. It is conceivable that the relationship between the Cartan theory and the non-abelian theory extends beyond the equality of their partition functions. For example, if we assume that both the non-abelian theory and the associated Cartan theory flow to SCFTs at low energies, the central charges are equal ---
\begin{equation}
\hat{c}_{C} = \sum_{i=1}^n \dim(R_i) (1-\rr_i) + (\dim(\lg) - \rank(\lg)) (-1) - \rank(\lg) = \hat{c}_{NA}\, .
\end{equation}
We leave the investigation of this relationship to future work.

\subsubsection*{Analytic structure of the integrand}

The integrand in equation \eqref{eq:ZS2} is a meromorphic function of the integration variables. We restrict ourselves to abelian gauge theories without loss of generality; when the GLSM is non-abelian, the analytic structure of the partition function is determined by the associated Cartan theory. The partition function for an abelian GLSM with $G=U(1)^s$ is
\begin{equation}
  \label{eq:abelianZS2}
  Z_{S^2} =  \int \left( \prod_{a=1}^{s} \frac{d \tau_a}{2 \pi i}\right)
  \,\,\, \sum_{\bm{m}\in \Z^s} \,\, e^{-4\pi \bm{r}\cdot \bt - i \bm{\theta} \cdot \bm{m}}\,\, \prod_{i=1}^n \,\,Z_{\phi_i}\, ,
\end{equation}
with
\begin{align}
  \label{eq:abelianZiZc}
Z_{\phi_i} = \frac{\Gamma(\frac{\rr_i}{2} - \bm{Q}_i \cdot(\bm{\tau} + \frac{\bm{m}}{2}) )}{\Gamma(1-\frac{\rr_i}{2} + \bm{Q}_i\cdot (\bm{\tau} - \frac{\bm{m}}{2}) )}.
\end{align}
Recall that $\Gamma(z)$ is a meromorphic function in the complex plane with simple poles at $z = -n$, $n\in \Z_{\geq 0}$, with residue
$\Res_{z=-n} \Gamma(z) = \frac{(-1)^n}{n!}\,$ and an essential singularity at $z=\infty$. Each factor $Z_{\phi_i}$ in the integrand has poles along the hyperplanes
\begin{equation}
  \label{eq:Zipoles}
H_i^{(k)}: \qquad  \frac{\rr_i}{2} - \bm{Q}_i \cdot(\bm{\tau} + \frac{\bm{m}}{2}) = -k\, , \quad k \in \Z_{\geq 0}, \quad k \geq \bm{Q}_i\cdot \bm{m}\, .
\end{equation}
The partition function integrand can be regarded as a meromorphic function on the space $\C^{\rank(\lg)}$ with poles along the hyperplanes $H_i^{(k)}$, which we refer to as \emph{polar divisors}. Similarly, each factor $Z_{\phi_i}$ has a zero along the hyperplanes
\begin{equation}
  \label{eq:Zizeros}
\mfH_i^{(k)}: \qquad 1-\frac{\rr_i}{2} + \bm{Q}_i\cdot (\bm{\tau} - \frac{\bm{m}}{2})    = -k\, , \quad k \in \Z_{\geq 0}, \quad k \geq \bm{Q}_i\cdot \bm{m}\, ,
\end{equation}
which we refer to as \emph{zero divisors}.
Note that the collection of hyperplanes $H_i^{(k)}$ is contained in the half space $\bm{Q}_i\cdot \bm{\tau} \geq 0$, while the $\mfH_i^{(k)}$ are contained in the half space $\bm{Q}_i\cdot \bm{\tau} \leq 0$. The analytic structure of the integrand will be relevant in what follows, as we will evaluate the integral in \eqref{eq:abelianZS2} through the method of residues.

\section{Phases of Gauged Linear Sigma Models}
\label{sec:phases}

The low-energy effective theory describing the dynamics of the GLSM
depends on the value of the FI parameters \cite{Witten:1993yc}. The space of FI
parameters can be divided into phase regions depending on the
character of the low-energy dynamics, and the analysis of the phase
structure of a GLSM typically proceeds through a close study of the
D-terms and F-terms. In this section we discuss how this information
can be easily extracted from $Z_{S^2}$. 

The idea is stated rather simply: when $Z_{S^2}$ is evaluated by the
method of residues, the contour prescription depends on the value of
the FI parameters, which in turn affects the set of poles that
contribute to the integral. At certain codimension-one walls in FI
parameter space the structure of poles contributing to the $Z_{S^2}$
integral can change, signaling the presence of a GLSM phase transition
along that wall. In particular, for abelian GLSMs we show that this recovers
the description of phases in terms of the ``secondary fan'' 
\cite{MR1020882,BFS,Aspinwall:1993nu}. Furthermore, we also demonstrate that phases
of non-abelian GLSMs can be understood in terms of phases of the
associated Cartan theory.

\subsection{Evaluation of the Partition Function}
\label{sec:Z-eval}

We use the multi-dimensional residue method (see \cite{GH}, for example) to evaluate the $s$-dimensional integral in \eqref{eq:abelianZS2} for an abelian GLSM as defined in section \ref{sec:two-sphere}. 
Let $I$ denote an $s$-element subset of $\{1, \cdots, n\}$ such that the $s$ vectors $\{ \bm{Q}_{i \in I} \}$ are linearly independent, and let $\mfI$ denote the set of all such subsets $I$. The corresponding polar divisors
$H^{(k)}_{i \in I}$ for all $k \in \Z_{\geq 0}$ 
(defined in \eqref{eq:Zipoles}),
 intersect in an infinite discrete point set that we denote by $P_I$. The residue of the integrand at any point $p \in P_I$ is well defined and can be evaluated. $Z_{S^2}$ is a sum of the residues at all points $p \in \bigcup_{I\in \mfI} P_I$ that lie inside the domain of integration.

The integration in \eqref{eq:abelianZS2} is carried out over the imaginary axis from $-i\infty$ to $+i\infty$ for each $\tau_a$. Each of the $s$ contours can be closed at infinity based on the asymptotic behavior of the integrand, which is entirely determined by the exponential factor, $e^{-4\pi \bm{r} \cdot \bt}$, for sufficiently large $|\bm{r}|$.\footnote{The condition $\sum_{i=1}^n \bm{Q}_i = 0$ ensures that the integrand behaves as an exponential at $\bm{\tau}=\infty$ in any wedge in $\C^s$.} Given a set $I$ as defined above, we express the FI parameters
as  $\bm{r} = \sum_{i\in I} r_i \bQ_i$. We have 
\begin{equation}
\Zc \sim e^{-4\pi r_i \bQ_i \cdot \tau}.
\end{equation}
If $r_i > 0$, for all $i \in I$, the points $P_I$ lie inside the domain of integration when the contours are closed at infinity, and $Z_{S^2}$ receives contributions from the residue at every point $p \in P_I$. This defines a cone in the space $\R^{s}$
\begin{equation}
C(I):=\left\{\bm{r} = \sum_{i\in I} r_i \bQ_i\,  \big\vert \ r_i > 0 \,\,\, \forall\  i \in I \right\} .
\end{equation}
The cones $C(I)$ for all $I \in \mfI$ overlap, and a \emph{generic} $\bm{r}$ lies in the interior of multiple cones. Let $\cC$ denote the (non-empty) intersection of all cones $C(I)$ that contain $r$. The partition function, for $\bm{r} \in \cC$, can be evaluated as
\begin{equation}
Z_{S^2} (\bm{r}) = \sum_{\tiny \begin{array}{c} I \in \mfI, \\ \bm{r} \in C(I)\end{array}} \sum_{p \in P_I} \Res_{\bm{\tau} = \bm{\tau}_p} \left( \sum_{\bm{m}\in \Z^s} \,\, e^{-4\pi \bm{r}\cdot \bt - i \bm{\theta} \cdot \bm{m}}\,\, \prod_{i=1}^n \,\, \frac{\Gamma(\frac{\rr_i}{2} - \bm{Q}_i \cdot(\bm{\tau} + \frac{\bm{m}}{2}) )}{\Gamma(1-\frac{\rr_i}{2} + \bm{Q}_i\cdot (\bm{\tau} - \frac{\bm{m}}{2}) )} \right)\, ,
\end{equation}
where $\bm{\tau}_p$ denotes the coordinates of the point $p \in P_I$. The above expression is an infinite series, whose convergence is controlled by the exponential factor $e^{-4\pi \bm{r}\cdot \bt - i \bm{\theta}\cdot \bm{m}}$, up to a finite shift of  $(\bm{r}, \bm{\theta})$ due to the exponential asymptotics of the remainder of the integrand. When $\bm{r}$ is sufficiently deep inside the cone $\cC$ all summations are convergent. At the cone boundary, convergence is not absolute but depends on the value of $\bm{\theta}$; for particular values of $\bm{\theta}$ this series is divergent.

We thus see that the expression \eqref{eq:abelianZS2} for $Z_{S^2}$ is an integral
of Mellin--Barnes type, which allows comparison of the behavior in different
regions of the FI space.  (See \cite{math.AG/9912109} for an early application
of this idea to GLSMs.)

\subsection{Abelian GLSM Phases and the Secondary Fan}
\label{sec:abelian phases}

We next study how the behavior of the two-sphere partition function changes as we tune the value of the FI parameter $\bm{r}$. 

Consider the set $\bm{C}_{max}$ of non-empty intersections, $\bigcap_{I\in S} C(I)$  for all subsets $S \subset \mfI $. The elements of $\bm{C}_{max}$ are a set of non-overlapping open cones that partition the space $\R^s$. The complement of $\bm{C}_{max}$, denoted by $\bm{B}$ is a closed subset of $\R^s$ and is a union of codimension-one hyperplanes, which we refer to as \emph{D-boundaries}, containing the origin. A generic $\bm{r}$ lies in the interior of exactly one element, say $\cC_1 \in \bm{C}_{max}$. As $\bm{r}$ is tuned it can cross a D-boundary and enter another cone $\cC_2 \in \bm{C}_{max}$. 

\begin{figure}
\centering
\begin{tikzpicture}
\draw[<->] (-3.2,0)--(3.2,0);
\draw[>->, thick] (0,-2)--(0,-1);
\draw[->, thick] (0,-1)--(0,0);
\draw[->, thick] (0,0)--(0,1);
\draw[->, thick] (0,1)--(0,2);
\fill[black]  (0.2,0) circle (0.07cm);
\fill[black]  (1.2,0) circle (0.07cm);
\fill[black]  (2.2,0) circle (0.07cm);
\draw  (-0.2,0) circle (0.07cm);
\draw  (-0.4,0) circle (0.07cm);
\draw  (-0.6,0) circle (0.07cm);
\draw  (-0.8,0) circle (0.07cm);
\draw  (-1.0,0) circle (0.07cm);
\draw  (-1.2,0) circle (0.07cm);
\draw  (-1.4,0) circle (0.07cm);
\draw  (-1.6,0) circle (0.07cm);
\draw  (-1.8,0) circle (0.07cm);
\draw  (-2.0,0) circle (0.07cm);
\draw  (-2.2,0) circle (0.07cm);
\draw  (-2.4,0) circle (0.07cm);
\draw  (-2.6,0) circle (0.07cm);
\draw  (-2.8,0) circle (0.07cm);
\draw  (-3.0,0) circle (0.07cm);
\end{tikzpicture}
\caption{Pole structure of the partition function integrand for the quintic Calabi--Yau threefold GLSM. The solid black circles are fourth order poles while the unfilled circles are simple poles. The integration path is indicated and is along the imaginary axis.} 
\label{fig:quintic-poles}
\end{figure}
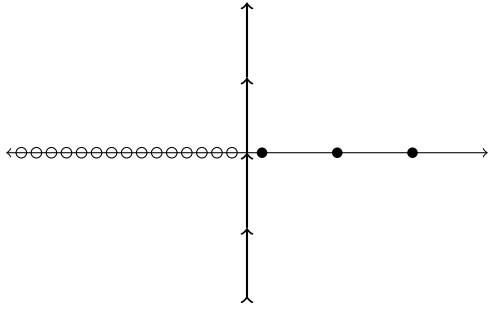

As a simple illustration, it is useful to consider
the case of a GLSM for a quintic hypersurface in $\P^4$. The GLSM has $G=U(1)$ and six fields $\Phi_{i=1,\cdots, 6}$ with charges $\{1,1,1,1,1,-5\}$. The partition function integrand has poles as shown in Figure \ref{fig:quintic-poles}. The maximal cones are $\cC_1 = \{ r>0\}$, $\cC_2 = \{r<0\}$, and the D-boundary is the point $r=0$. When $r>0$ we close the contour in the right half plane, where $Z_{S^2}$ receives contributions from the fourth-order poles, and when $r < 0$, $Z_{S^2}$ receives contributions from the simple poles in the left half plane.

The set of cones $\cC \in \bm{C}_{max}$ are familiar in the context of toric geometry: they are the maximal cones in
the secondary fan.  Note that there are multiple descriptions of the secondary fan, as
mentioned in \cite{Stienstra:2005nr}, for example. Here we have
recovered the \emph{Gale Transform} description of the secondary fan,
which utilizes GLSM charge data directly. Specifically, it is a
complete fan in the space of FI parameters, whose one-dimensional cones are defined by the GLSM charges $\bQ_{i=1, \cdots, n}$.
The vectors $\bQ_i$ also generate the maximal cones
$\cC$ defined above and it is a straightforward, though sometimes
tedious, calculation to determine $\cC$ explicitly.
The maximal cones of the secondary fan determine the D-term phase
structure of abelian GLSMs \cite{Aspinwall:1993nu}. This structure is encoded in the partition function in
terms of the residue sets $P_I$ that are contained within the domain of integration specified by $\bm{r}$.

In crossing from maximal cone $\cC_1$ to $\cC_2$, $\bm{r}$ must enter or exit some $C(I)$, and therefore, a set of residues associated to the corresponding $P_I$ will either enter or exit the domain of integration respectively. The residue contribution from a set $P_I$ can be zero if sufficiently many zero divisors $\mfH_i^{(k)}$ pass through every point in $P_I$. When this is the case the infinite series defining the partition function does not diverge at the boundary $B$ between $\cC_1$ and $\cC_2$. Effectively, this erases the boundary $B$ and enlarges the maximal cone $\cC_1$ to $\cC_1 \cup \cC_2$. 
We will refer to D-boundaries associated with sets $P_I$ --- whose residues are non zero --- as \emph{Z-boundaries}. 
This phenomenon is familiar in the study of GLSM phases. The vacuum moduli space of a GLSM can behave smoothly across a D-boundary defined by the D-terms, since it is a solution of both the D-term and the F-term equations. An example of this phenomenon occurs in certain Calabi--Yau hypersurfaces in toric varieties where a curve that is not contained in the hypersurface can be flopped. The ambient space goes through a flop while the hypersurface is unaffected. We will encounter and exploit this phenomenon is section \ref{sec:P2}.

The true phase boundaries of the GLSM can be obtained by an analysis of both the D-terms and F-terms. In a number of examples, we have found that these true phase boundaries agree with the Z-boundaries defined by the partition function. However, it is conceivable that there could be phase boundaries of a GLSM which are unseen by the two-sphere partition function, although other observables suffer discontinuities. We leave the investigation of this possibility, and the relationship of the true phase boundaries and the Z-boundaries, to future work.

\subsection{Non-Abelian GLSM Phases}

To evaluate the partition function for non-abelian GLSMs we can compute
the partition function of the (abelian) associated Cartan theory using
the approach described above. The extra FI parameters $\tilde{r}_\mu$,
$\mu = s+1, \cdots, \rank(\lg)$, in the Cartan theory are useful as
regulators in evaluating the necessary contour integrals. 

The partition function for the Cartan theory is given by
\begin{eqnarray}
Z_{S^2}^C(r_\mu, \theta_\mu) = \frac{1}{|\mathcal{W}|} \sum_{\bm{m} \in \Z^{\rank(\lg)}}\ \  \int \ \left( \  \prod_{\mu=1}^{\rank(\lg)} \frac{d \tau_\mu}{2 \pi i}
 e^{-4\pi  r_\mu \tau_\mu - i  \theta_\mu m_\mu}  \ \right) \nonumber \\
 \prod_{\alpha \in \mathcal{P}} \left( 
  \frac{(\alpha_\mu { m}_\mu)^2}{4} - 
  (\alpha_\mu  \tau_\mu)^2\right)\, 
 \prod_{\rho^i \in R_i} \frac{\Gamma(\frac{\rr_i}{2}  - \rho^i_\mu ( \tau_\mu + \frac{ m_\mu}{2}))}{\Gamma(1-\frac{
\rr_i}{2} +\rho^i_\mu  ( \tau_\mu - \frac{ m_\mu}{2}))}\, ,
\end{eqnarray}
where we have dropped the \~{ } on the extra FI parameters $\tilde r_\mu$, $\mu = s+1, \cdots, \rank(\lg)$.
The Weyl group $\cW$ is generated by reflections $g_\beta$ about the planes perpendicular to the simple roots $\beta$ of the Lie algebra. Each $g_\beta$ has an action on the space $\R^{\rank(\lg)}$ and is represented by an orthogonal matrix $M_\beta$ \cite{cahn1984semi}. Since the group elements $g_\beta$ permute the weights, and take all the positive roots to themselves, except for the root $\beta$ (which is mapped to $-\beta$), the following identity holds
\begin{equation}
Z_{S^2}^C (r_\mu (M_\beta)_{\mu\nu}, \theta_\mu (M_\beta)_{\mu\nu}) = Z_{S^2}^C (r_\mu,\theta_\mu)\, .
\end{equation}
Since the elements $g_\beta$, where $\beta$ is a simple root, generate $\cW$, the above identity holds for all orthogonal matrices $M_g$, for $g \in \cW$. Therefore, the phase diagram of the Cartan theory (inferred from the partition function) is symmetric under the action of the Weyl group.

The FI parameter space $\R^s$ of the non-abelian theory is the Weyl-invariant subspace of the FI parameter space $\R^{\rank(\lg)}$ of the Cartan theory. Any phase boundary in the non-abelian theory across which the partition function $Z_{S^2}^{NA}$ is divergent (for some values of $\bm{\theta}$) must also correspond to a phase boundary in the Cartan theory. The phase boundaries of the non-abelian theory are obtained by restricting the phase diagram of the Cartan theory -- and therefore its secondary fan -- to its Weyl-invariant subspace. We now apply this method to two examples of non-abelian GLSMs that have been studied in the literature.

\noindent
{\bf Pfaffian GLSM:}
This GLSM was studied in \cite{Hori:2006dk} and has gauge group $U(2)$. The matter content of the non-abelian theory and the Cartan theory are described below:

\begin{minipage}{0.4\textwidth}
\begin{center}
Non-abelian theory \\[0.1in]

\begin{tabular}{ccc}
 & $U(2)$ & $U(1)_v$ \\
\hline
$\Phi_{i=1, \cdots, 7}$ & $\Box$ & $2\rr_\phi$ \\
$P_{a=1,\cdots, 7}$ & $\det^{-1}$ & $2-2\rr_\phi$ \\
\hline \\
\end{tabular}
\end{center}
\end{minipage}
\begin{minipage}{0.4\textwidth}
\begin{center}
Associated Cartan theory \\[0.1in]

\begin{tabular}{cccc}
 & $U(1)_1$ & $U(1)_2$ & $U(1)_v$ \\
\hline
$\Phi^1_{i=1, \cdots, 7}$ & 1 & 0 & $2\rr_x$ \\
$\Phi^2_{i=1, \cdots, 7}$ & 0 & 1 & $2\rr_x$ \\
$\Phi_{a=1,\cdots, 7}$ & $-1$ & $-1$  & $2-2\rr_x$ \\
$A_\pm$ & $\pm 1$ & $\mp 1$ & $2$ \\
\hline \\
\end{tabular}
\end{center}
\end{minipage}

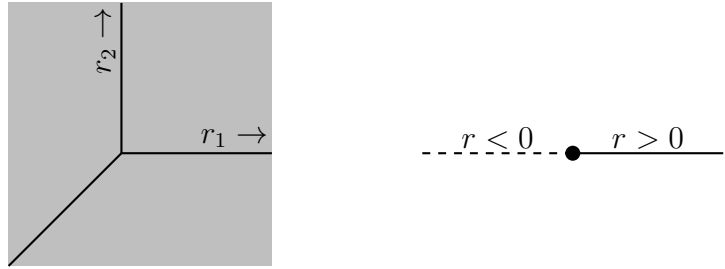
\begin{figure}[b]
\centering
\begin{tikzpicture}
\fill[lightgray] (-1.5,-1.5)--(-1.5,2)--(2,2)--(2,-1.5)--(-1.5,-1.5);
\draw[thick] (0,0)--(2,0);
\draw[thick] (0,0)--(0,2);
\draw[thick] (0,0)--(-1.5,-1.5);
\draw[thick,dashed] (4,0)--(6,0);
\draw[thick] (6,0)--(8,0);
\fill[black] (6,0) circle (.1cm);
\node at (5,.2) [] {$ r<0$};
\node at (7,.2) [] {$r >0$};
\node at (1.5,.2) [] {$r_1 \rightarrow$};
\node[rotate=90] at (-0.2,1.5) [] { $r_2 \rightarrow$};
\end{tikzpicture}
\caption{Phase diagram of the Pfaffian GLSM obtained by projection from the Cartan theory.}
\label{fig:pfaff}
\end{figure}

The phase diagram of the non-abelian GLSM is obtained by restricting the phase diagram of the Cartan theory to the subspace $r_1=r_2$ as shown in Figure \ref{fig:pfaff}. Note that the $r<0$ phase of the non-abelian theory is actually a phase boundary in the associated Cartan theory. This is consistent with the fact that the restriction to $r_1=r_2$ puts the infinite sum over residues at its radius of convergence as noted in \cite{Jockers:2012dk}.

\noindent
{\bf Gulliksen-Neg\r{a}rd GLSM: } The Gulliksen-Neg\r{a}rd (GN) GLSM with gauge group $U(2)\times U(1)$ was studied in 
\cite{Jockers:2012dk}. The matter content of the non-abelian and associated Cartan theories are listed below:

\begin{minipage}{0.32\textwidth}
\begin{center}
Non-abelian theory \\[0.1in]

\begin{tabular}{cccc}
 & $U(1)_0$ & $U(2)$ & $U(1)_v$ \\
\hline
$\Phi_{a=1, \cdots, 8}$ & 1 & $1$ & $2\rr_\phi$ \\
$X_{i=1,\cdots, 4}$ & $0$ & $\overline{\Box}$ & $2\rr_x$ \\
$P_{i=1,\cdots, 4}$ & $-1$ & $\Box$ & $2-2\rr_x-2\rr_\phi$ \\
\hline
\end{tabular}
\end{center}
\end{minipage}
\begin{minipage}{0.6\textwidth}
\begin{center}
Associated Cartan theory \\[0.1in]

\begin{tabular}{ccccc}
Field & $U(1)_0$ & $U(1)_1$ & $U(1)_2$ & $U(1)_v$  \\ \hline
$\Phi$ & $1$ & $0$ & $0$ & $2\rr_\phi$  \\
$X_1$ & $0$ & $-1$ & $0$ & $2\rr_x$  \\
$X_2$ & $0$ & $0$ & $-1$ & $2\rr_x$  \\
$P_1$ & $-1$ & $1$ & $0$ & $2-2\rr_x-2\rr_\phi$  \\
$P_2$ & $-1$ & $0$ & $1$ & $2-2\rr_x-2\rr_\phi$ \\
$A_\pm$ & $0$ & $\pm 1$ & $\mp 1$ & $2$ \\
\hline \\
\end{tabular}
\end{center}
\end{minipage}

\begin{figure}[t]
 \centering
 \includegraphics[width=1.5in]{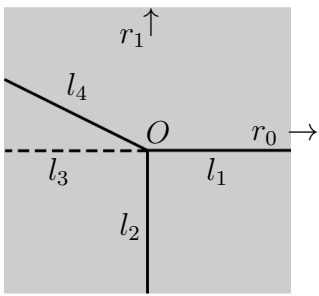}
 \put(-32,43){$l_1$}
 \put(-92,43){$l_3$}
 \put(-65,22){$l_2$}
 \put(-85,75){$l_4$}
 \put(-15,58){$r_0 \rightarrow$}
 \put(-65,95){$r_1$}
\put(-55,57){$O$}
 \put(-56,100){$\uparrow$}
 \caption{The candidate phase boundaries in the
   Gulliksen-Neg\r{a}rd GLSM obtained by projection from the associated Cartan theory. The boundary $l_3$ is dashed to indicate that it is actually lifted in the IR SCFT.}
 \label{fig:phases-classical}
 \end{figure}

Phase boundaries associated with changes in $Z_{S^2}^{NA}$ can be
determined by comparison to the secondary fan of $Z_{S^2}^C$. 
We carry out the secondary fan analysis using only the charge vectors $\bQ_{\Phi}$, $\bQ_{X_1}$, $\bQ_{X_2}$, $\bQ_{P_1}$, $\bQ_{P_2}$, since the one-loop determinants associated to $A_\pm$ do not contribute any poles.
Maximal cones in
the secondary fan are three-dimensional, and entering a different
maximal cone by varying FI parameters requires going through a face of
some cone $C(I)$. 
This face is itself a cone $C(I')$, where
$I'=\{\bQ_i,\bQ_j\}$ for two of the $\bQ$-vectors above. 
Intersecting these cones with the Weyl-invariant subspace $r_1=r_2$
yields four lines and one point in $(r_0,r_1)$-space depicted in Figure \ref{fig:phases-classical}. They are given by
\begin{table}[h]
\centering
\begin{tabular}{c|c|c}
locus $C(I') \cap \{r_1=r_2 \}$ & Different Sets $I'$ & Label \\ \hline
$r_0 \ge 0$, $r_1 =0$ & $\{\bQ_{X_2},\bQ_{\Phi}\}$, $\{\bQ_{X_1},\bQ_{\Phi}\}$ & line $l_1$\\
$r_0 = 0$, $r_1 \le 0$ & $\{\bQ_{X_2},\bQ_{X_1}\}$ & line $l_2$\\
$r_0 \le 0$, $r_1=0$ & $\{\bQ_{X_1},\bQ_{P_1}\}$$\{\bQ_{X_2},\bQ_{P_2}\}$ &line  $l_3$\\
$r_0 + 2r_1=0$, $r_0\leq 0$ & $\{\bQ_{P_2},\bQ_{P_1}\}$ &line  $l_4$\\
$r_0=r_1=0$ & $\{\bQ_{X_1},\bQ_{P_2}\}$,$\{\bQ_{X_2},\bQ_{P_1}\}$, $\{\bQ_{P_1},\bQ_{\Phi}\}$, $\{\bQ_{P_2},\bQ_{\Phi}\}$ & pt $O$\\
\end{tabular}
\end{table}

The phase structure of this GLSM was analyzed in \cite{Jockers:2012zr} through a
D-term and F-term analysis and revealed a phase diagram with only
three boundaries --- $l_1, l_2, l_4$. This is not an inconsistency
(see discussion in \ref{sec:abelian phases}) as there could be
accidental cancellations of residues resulting in certain boundaries
being deleted from the phase structure. In related GLSMs with gauge group $U(k) \times U(1)$, studied in \cite{Jockers:2012zr}, the boundary $l_3$ is a true phase boundary; determining whether this is the case
requires a more detailed analysis. Nevertheless, we have seen that the secondary fan
of the Cartan theory provides valuable information about phase boundaries
in the Gulliksen-Neg\r{a}rd GLSM.

\section{Geometric Phases of GLSMs}
\label{sec:geometries}

Having discussed phase transitions, in this section we address the
physics of particular phases.  We describe a set of conditions on
$Z_{S^2}$ which are necessary for a given GLSM phase to be geometric,
\ie, the physics in the IR is equivalent to that of a smooth
Calabi--Yau NLSM. Typically, a careful D-term and F-term analysis is
necessary to ascertain whether or not a particular GLSM phase is
geometric; such a phase analysis can be quite involved as illustrated
by the analysis in \cite{Hori:2006dk}. The conditions we introduce
give powerful checks on such analyses.

\subsection{Criteria for Identifying Geometric Phases}
\label{sec:criteria}
Let us discuss criteria for geometric phases.
Consider a GLSM that flows to an $(2,2)$ SCFT with central charge $\hat{c}  = 3$. It satisfies \cite{Witten:1993yc, Silverstein:1994ih}
\begin{align}
\text{\bf Condition one}&: \qquad \qquad \sum_{i=1}^n \tr_{R_i} (\mathfrak{t})  =  0 \, , \quad \mathfrak{t} \in \mathfrak{g}\, , \label{eq:MUM1} \\
\text{\bf Condition two}&: \qquad \qquad\sum_{i=1}^n (1-\rr_i)\, \dim(R_i) - \dim(\lg)  = \hat{c}=3\, . \label{eq:MUM2}
\end{align}
The structure of the chiral ring of SCFTs associated with smooth Calabi--Yau threefolds requires \cite{Lerche:1989uy} that gauge invariant operators have
integral left and right moving R-charges. This gives
\begin{equation}
\mbox{{\bf Condition three}: the vector R-charge of every gauge-invariant chiral local operator} \in 2\bZ \label{eq:MUM3}\, ,
\end{equation}
 since the chiral fields have
zero axial R-charge.

Consider a phase of this GLSM defined by a region $\sum_{b=1}^s \ p_{ab} r_b > 0$ for all $a\,$. This phase corresponds to a singular point in the complexified FI parameter space located at $\xi_a=0$, $a=1, \cdots, s$, where $\xi_a:= \prod_{b=1}^s\ z_b^{p_{ab}}$ are local coordinates in the punctured neighborhood of the singular point.
If we assume that the GLSM has a geometric phase which corresponds to a smooth Calabi--Yau threefold $Y$ (the point at $\xi_a=0$ is then called a \emph{large volume point}), then the classical metric on the K\"ahler moduli space is specified through the following K\"ahler potential \cite{Candelas:1990pi}
\begin{equation}
e^{-K}\big|_{\mbox{classical}} = \frac{1}{6} \int_Y J \wedge J \wedge J = \frac{1}{6} \kappa_{abc} J_a J_b J_c\, .
\end{equation}
Here $J = \sum_{a=1}^s J_a \omega_a$ is the K\"ahler form on $Y$ that we have expanded in terms of the integral basis $\omega_a$ of $H^{1,1}(Y,\Z)$, which defines the intersection form $\kappa_{abc}:=\int_Y \omega_a\wedge \omega_b\wedge \omega_c$. Therefore, the two-sphere partition function computed in this phase must have the following asymptotic behavior in the $\xi_a \rightarrow 0$ limit
\begin{equation}
\lim_{\xi \rightarrow 0} \ Z_{S^2}(\xi, \xib) \ = \ \frac{1}{6} \sum_{a,b,c =1}^s \kappa_{abc} \log(\xi_a\xib_a) \log(\xi_b\xib_b) \log(\xi_c\xib_c)\, , \label{eq:Z-asymp}
\end{equation}
since the K\"ahler class and FI parameters are related through $J_a \propto \log(\xi_a\xib_a)$ close to a large volume point \cite{Witten:1993yc}. Note that we have absorbed any positive numerical constants in \eqref{eq:Z-asymp} using a K\"ahler transformation. This leads us to
\begin{equation}
\mbox{{\bf Condition four}: $Z_{S^2}$ must have the asymptotics as in equation \eqref{eq:Z-asymp}}\label{eq:MUM4}
\end{equation}
which is a necessary condition for the existence of a geometric phase.

Let us briefly discuss some relevant background before stating our final criterion necessary for the
existence of a geometric phase.
Any $\hat{c}=3$ SCFT satisfying condition \eqref{eq:MUM3} can be used as the worldsheet theory of the type II superstring to generate a four-dimensional $\cN=2$ supergravity \cite{Sen:1986mg, Seiberg:1988pf} .  The $s$-dimensional complexified FI parameter space $\cM$ is governed by \emph{special geometry} \cite{deWit:1984pk, Periwal:1989mx, Strominger:1990pd, Candelas:1990pi, Craps:1997gp, MR717607, Freed:1997dp}. Let $(\xi_a)$, $a = 1, \cdots, s$  be local coordinates on $\cM$. 
Special geometry implies that $\cM$ is a Hodge--K\"ahler manifold that carries 
an $Sp(2s+2,\Z)$  bundle with local sections $X^I(\xi)$, $I = 0, \cdots, s$. In addition, there exists a holomorphic function $\cF(X)$, known as the \emph{prepotential} that is homogeneous and has degree 2. Locally, $(X^I, \frac{\partial \cF}{\partial X^I})$ specify a choice of frame for the $Sp(2s+2,\Z)$ bundle and are known as the \emph{periods}. The K\"ahler potential on moduli space is specified through the periods as
\begin{equation}
e^{-K(\xi,\xib)} = i\left(X^I \overline{\frac{\partial \cF}{\partial X^I}} - \overline{X^I} \frac{\partial \cF}{\partial X^I} \right)\, . \label{eq:Z-periods}
\end{equation}  
For definitions and further details about special geometry we refer the reader to \cite{Craps:1997gp, Freed:1997dp}.

When the GLSM is in a geometric phase
the periods $(X^I, \partial_I \cF)$ have a special structure: in the $\xi_a \rightarrow 0$ limit, the leading behavior is given by
\begin{gather}
X^0 \sim \prod_{a=1}^s \xi_a^{\delta_a} \, , \quad X^{b} \sim X^0(\xi) \log \xi_b \, , \quad \partial_{b} F \sim X^0(\xi) C_{abc} \log \xi_b \log \xi_c\, , \nonumber \\
 \partial_0 F \sim X^0(\xi) C_{abc} \log \xi_a \log \xi_b \log\xi_c \, , \quad b=1, \cdots, s\, , \label{eq:logperiods}
\end{gather}
where $\delta_a \in \mathbb{Q}$. This behavior of the periods as $\xi \rightarrow 0$ is usually referred to as \emph{maximally unipotent monodromy (MUM)}, which is a statement about the form of the $Sp(2s+2,\Z)$ monodromy around any boundary divisor passing through $\xi=0$ \cite{mirrorguide,compact}.

Recall the evaluation of the partition function by the method of residues discussed in Section \ref{sec:Z-eval}. 
Consider all $rk(\lg)$-element subsets $I \subset \{1, \cdots, n\}$ such that the corresponding pole hyperplanes intersect in a point set $P_I$. Let $\cI_0$ denote the set of all subsets $I$ whose point sets $P_I$ lie inside the contours of integration defined by the GLSM phase. 
The partition function is evaluated by adding up all the residues at the points  in $\bigcup_{I\in \cI_0} P_I$. If $P_I \neq P_J$ for some $I,J \in \cI_0$ the contributions to the partition function from the two sets of residues will involve distinct power series expansions. 
We suspect that this signals the presence of at least two independent periods, which in the $\xi \rightarrow 0$ limit do not diverge as a power of $\log \xi_a$. If this is the case, the periods do not have the $\log$-structure  as in equation \eqref{eq:logperiods} that is characteristic of MUM. We are led to a condition
\begin{equation}
P_I = P_J\, \quad \forall \ I, J \in \cI_0\, . \label{eq:condition5}
\end{equation}
We would like to stress that we do not have a proof that this condition on the residue structure of the partition function is necessary for MUM. We will return to this condition in the context of a particular example in Section \ref{sec:resolution}.

We have discussed four necessary conditions for the existence of a geometric phase in a GLSM. We  refer to these as the \emph{geometric phase criteria} and for convenience list them in Table \ref{tab:gpc}.
\begin{table}
\begin{tabular}{|l|p{4.8in}|}
\hline
1. & $\sum_{i=1}^n \tr_{R_i} (\mathfrak{t})  =  0 \, , \quad \mathfrak{t} \in \mathfrak{g}\, .$  \\[0.1in]
2. & $\sum_{i=1}^n (1-\rr_i)\,\dim(R_i) - \dim(\lg)  = \hat{c}=3\, . $ \\[0.1in]
3. &  The vector R-charge of every gauge-invariant chiral local operator is an even integer. \\[0.1in]
4. &  $\lim_{\xi \rightarrow 0} \ Z_{S^2}(\xi, \xib) \ = \ \frac{1}{6} \sum_{a,b,c =1}^s \kappa_{abc} \log(\xi_a\xib_a) \log(\xi_b\xib_b) \log(\xi_c\xib_c)$. \\
\hline
\end{tabular}
\centering
\caption{The geometric phase criteria.} \label{tab:gpc}
\end{table}

\subsubsection*{Geometric Phase Criteria for abelian GLSMs}

For an abelian GLSM one of these criteria is equivalent to a
constraint on the partition function integrand which is easy to
check. Recall that the partition function for an abelian GLSM is given
in equation (\ref{eq:abelianZS2}).  
Each one-loop matter determinant
$Z_{\phi_i}$ 
has a pole along the hyperplanes  $H_i^{(k)}$ (polar divisors) and a zero along the hyperplanes  $\mfH_i^{(k)}$ (zero divisors).
Consider a point $p\in P_I$, for some $I\in \cI_0$, located at the codimension-$s$
intersection of the polar divisors $H_i^{(k_i)}$ for some $k_i$ and
all $i \in I$. 
Let $S_p$ and $S_z$ be the total number
of polar divisors and zero divisors containing $p$, respectively.
Then we define the \emph{multiplicity} $\mm$ of $p$ to be
$\mm=S_p - S_z - s + 1.$ When $\mm > 1$ we expand the integrand in a
multidimensional Laurent series and read off the residue as the
coefficient of the $\prod_{a=1}^s \frac{1}{\tau_a-\tau_a^{(p)}}$ term,
where $\bm{\tau}^{(p)}$ are the coordinates of $p$. In doing so we
have to take $\mm-1$ derivatives of the integrand in
\eqref{eq:abelianZS2}. Each derivative $\partial/\partial \tau_a$ results
in a $\log z_a$ term in the residue.  As a simple illustration of the
above phenomenon, when $s=1$ the multiplicity at $p$ is simply the
order of the pole at $p$. For a pole of order $\mm$,
\begin{equation}
\Res_{\tau=0} \ \frac{f(\tau)}{\tau^{\mm+1}} = \frac{1}{\mm !} \left. \frac{d^{\mm}f}{d\tau^{\mm}}\right|_{\tau=0}\, .
\end{equation}
If $f(\tau)$ contained a factor of $z^\tau$ the derivatives above would generate a $(\log z)^{\mm-1}$ term.

Given this relationship between $\mm$ and the log structure of the partition function, it is easy
to see that for an abelian GLSM criterion $4$ in
Table \ref{tab:gpc} is equivalent to the condition: 
\begin{align*}
\text{there exists a point $p \in P_I$ for some $I$ that has multiplicity $\mm=4$},
\end{align*}
which is very easy to check in examples.

\subsection{Are the criteria sufficient?}

We present a simple example that shows that the criteria in Table \ref{tab:gpc} are necessary but not sufficient conditions for the existence of a geometric phase. Consider a GLSM with gauge group $U(1)\times \Z_5$ and matter content as shown:
\begin{center}
\begin{tabular}{|c|ccccc|}
\hline
Field    & $X_{1,2,3,4}$ & $P_1$ & $P_2$ & $P_3$ & $P_4$\\
\hline
$U(1)$ &  $+1$   & $-1$   & $-1$     & $-1$ & $-1$ \\
$U(1)_R$& 0      &   $\frac{2}{5}$        &    $\frac{4}{5}$ & $\frac{6}{5}$  & $\frac{8}{5}$\\
\hline
\end{tabular}
\end{center}
The $\Z_5$ gauge symmetry acts as $P_j \rightarrow e^{2\pi i j/5} P_j$, $j=1,\cdots,4$. The most general superpotential is 
\begin{equation}
W = P_2 P_3 G_1(X)+P_1 P_4 G_2(X)+P_1 P_2^2 G_3(X)+P_1^2 P_3 G_4(X)+P_1^3 P_2 G_5(X)+P_1^5 G_6(X)
\, ,
\end{equation}
where $G_i(X)$ are homogeneous polynomials whose degrees are fixed by gauge invariance. 

The two-sphere partition function in the $r>0$ phase is 
\begin{equation}
Z_{S^2} = \sum_{n,l=0}^\infty \oint \frac{d\epsilon}{2\pi i} \ \frac{\pi^4}{\sin^4\pi \epsilon} \  z^{\ n-\epsilon} \zb^{\ l-\epsilon} \ \prod_{k=1}^4 \frac{\Gamma(\frac{k}{5} + n-\epsilon)\Gamma(\frac{k}{5} + l-\epsilon)}{\Gamma(1+n-\epsilon)\Gamma(1+l-\epsilon)}    \, .
\end{equation}
In fact, via $\Gamma$-function identities the expression above is equal to the partition function of a GLSM for a quintic hypersurface in $\P^4$ that was discussed briefly in Section \ref{sec:abelian phases}. The partition function only has poles of fourth order and all the geometric phase criteria are met in this GLSM. 

The vacuum moduli space of the GLSM is a solution to the D-term and F-term equations:
\begin{equation*}
\begin{array}{c}
 |x_1|^2+|x_2|^2 - |p_1|^2-|p_2|^2 = r\, , \\[0.1in]
 p_1 G_2(x) = 0 \, ,  \quad p_2 G_1(x)+p_1^2 G_4(x) = 0\, , \quad  p_3 G_1(x)+2 p_1 p_2 G_3(x)+p_1^3 G_5(x) = 0\\[0.02in]
 p_4 G_2(x)+p_2^2 G_3(x)+2 p_1 p_3 G_4(x)+3 p_1^2 p_2 G_5(x)+5 p_1^4 G_6(x) = 0\\[0.02in]
 p_2 p_3 \partial_iG_1(x)+p_1 p_4 \partial_iG_2(x)+p_1 p_2^2 \partial_iG_3(x)+p_1^2 p_3 \partial_iG_4(x)+p_1^3 p_2 \partial_iG_5(x)+p_1^5 \partial_iG_6(x)= 0
\end{array}
\end{equation*}
When $r>0$ we must have $x \neq 0$. It is easy to see that the classical vacuum moduli space has a non-compact branch of solutions:
\begin{equation}
p_1 =0\, , \quad p_2=0\, , \quad G_1(x)=0\, , \quad G_2(x)=0\, , \quad p_3, p_4 \mbox{ arbitrary.}
\end{equation}
Consequently, the low-energy field theory is singular and cannot be
identified with a compact Calabi--Yau NLSM, even though all the
geometric phase criteria were satisfied. It is intriguing that in
spite of this singular behavior, the procedure outlined in
\cite{Jockers:2012dk} can be used to extract the Gromov-Witten
invariants, which agree with those of the quintic Calabi--Yau
threefold. This example illustrates that the criteria are not
sufficient to ensure a geometric interpretation of the low-energy
physics as a NLSM with a compact Calabi--Yau threefold as target
space.

\subsection{The Geometric Criteria and Singular Geometries}
\label{sec:resolution}

In this section we study the geometric phase criteria  in GLSM phases, which correspond to singular limits of Calabi--Yau NLSMs. These serve as examples of GLSM phases that satisfy the conditions in Table \ref{tab:gpc}, but fail criterion \eqref{eq:condition5}.

\begin{table}
\centering
\begin{tabular}{|c|cccccc|}
\hline
field & $\phi_1$ &$\phi_2$ &$\phi_3$ &$\phi_4$ &$\phi_5$ & $P$ \\ \hline
$U(1)$ & $1$ & $w_2$ & $w_3$ & $w_4$ & $w_5$ & $-1 - \sum_{i=2}^5 w_i$ \\  
$U(1)_R$ & $0$& $0$& $0$& $0$& $0$& $2$ \\
\hline
 \end{tabular}
\caption{This table lists the field content of GLSMs that model a degree $\sum_{i=1}^5 w_i$ hypersurface in the weighted projective space $\P^{w_1, \cdots, w_5}$ .}
\label{table:singular GLSMs}
\end{table}

Consider the class of GLSMs defined by the field content in Table \ref{table:singular GLSMs} 
and the superpotential $W=PG(\phi)$, where $G$ is a quasi-homogeneous
polynomial of degree $ \sum_{i=1}^5 w_i$. 
The models with $(w_1, w_2,w_3,w_4,w_5)$
given by $(1,1,1,1,2)$, $(1,1,1,1,4)$ and $(1,1,1,2,5)$ 
have singularities at points in the
ambient weighted project space that miss the generic
 hypersurface $G(\phi)=0$. Therefore, these models have a geometric phase that can be interpreted as a NLSM with a smooth Calabi--Yau threefold as target space. All the criteria in Table \ref{tab:gpc} are satisfied in these examples.

\begin{table}
\centering
\begin{tabular}{|c|c|c|c|}
\hline
$(w_1,w_2,w_3,w_4,w_5 \ | \ -\sum_{i=1}^5 w_i)$ & $h^{1,1}$  & Distinct $P_I$ & $2h^{1,1}+2=\sum_{p\in \{box\}} \mm_{p}$ \\
\hline
$(1,1,2,2,2\ | \ -8)$ & 2 & 2 & 4+2\\
$(1,1,2,2,6\ | \ -12)$ & 2& 2 & 4+2\\
$(1,2,2,3,4\ | \ -12)$ & 2 & 2 & 4+2\\ 
$(1,2,2,2,7\ | \ -14)$ & 2& 2 & 4+2\\ 
$(1,1,1,6,9\ | \ -18)$ & 2 & 2 & 4+1+1\\
\hline
$(1,1,1,3,6\ | \ -12)$ & 3& 3 & 4+2+2\\
$(1,2,3,3,3\ | \ -12)$ & 3& 3 & 4+2+2\\
$(1,3,3,3,5\ | \ -15)$ & 3& 3& 4+2+2\\
$(1,2,3,3,9\ | \ -18)$ & 3& 3& 4+2+2\\
$(1,1,2,8,12\ | \ -24)$ &3 & 3& 4+1+2+1\\
\hline
\end{tabular}
\caption{The first column defines the GLSM through its $U(1)$-charges. The second column is the number of K\"ahler moduli in the resolved geometry. The third column denotes the number of distinct $P_I$ whose residues are non-zero and that contribute to the contour integral. The fourth column demonstrates the relationship between pole multiplicities and $2h^{1,1}+2$.}
\label{table:resolution}
\end{table}

The ambient space singularities for the the models given in Table
\ref{table:resolution} have singularities on curves and/or surfaces
that intersect the generic hypersurface. These singularities can be
resolved through blowups of the ambient space and result in smooth
Calabi--Yau threefolds with two or three moduli 
\cite{Candelas:1993dm,Hosono:1993qy,Candelas:1994hw}. The $r > 0$
phase of these GLSMs satisfies the first four criteria listed in Table
\ref{tab:gpc}, in particular, the partition function has the right
asymptotics as $z \rightarrow 0$. In each case, however, there are
non-zero residue contributions to $Z_{S^2}$ arising from multiple,
distinct sets $P_I$. Interestingly, the number of such distinct $P_I$
exactly equals the number of K\"ahler moduli in the resolved
threefold. It is tempting to speculate that a GLSM phase that
satisfies the geometric phase criteria, but not condition \eqref{eq:condition5}, may signal a singular
geometry of the kind encountered above. 

Moreover, a careful analysis of the pole structure using the method of
appendix \ref{sec:comments and grid} reveals that, within a given
modular box in the singular GLSM partition function, the sum of the
multiplicities $\mm$ for contributing poles is precisely $2 h^{1,1} +
2$, where $h^{1,1}$ is the number of K\" ahler moduli in the resolved
geometry. This matches the $2h^{1,1}+2$ periods that enter the
partition function in the resolved GLSM, and the fact that the
counting matches the unresolved GLSM suggests that the partition
function captures a singular limit of the  periods.

\section{Illustrative Examples}
\label{sec:illustrative examples}
In this section we apply the geometric phase criteria in Table
\ref{tab:gpc} and identify novel geometric phases of GLSMs.  We study
a collection of GLSMs with gauge group $U(1)$ and charged matter fields
that satisfy the above criteria. We focus our attention on one member
of the collection, which we call the $P^2$-GLSM, and perform an analysis
of the D-term and the F-terms. The phase that satisfies the geometric
criteria appears as a hybrid phase $Y$, which can be visualized 
semi-classically as two copies of $\P^3$
glued together by a Landau--Ginzburg model. We provide a
physical argument that shows that this hybrid phase can be identified with an
NLSM with target space $X$, the double-cover of $\P^3$ branched over a
smooth octic surface. The argument involves studying the phase diagram of an
auxiliary GLSM, which we call GLSM2, with gauge group $U(1)\times
U(1)$. GLSM2 has three phases: the hybrid phase $Y$, a geometric phase
corresponding to the Calabi--Yau threefold $X$ and a Landau-Ginzburg
orbifold phase. We show that the IR physics of GLSM2 is identical
along a codimension 1 locus in the FI parameter space, and this locus
connects the singular points corresponding to the geometric phase $X$
and the hybrid phase $Y$. We believe this to be a strong physical
argument for a geometric interpretation of the hybrid phase of the
$P^2$-GLSM and the other GLSMs in the collection.


\subsection{Calabi--Yau Geometries from Gauge Theories}
\label{sec:P2}

We consider a special class of GLSMs with no $U(1)_A$ anomaly that flow to $(2,2)$ SCFTs with $\hat{c} = 3$ and integral R-charges. 
The gauge group is $G = U(1) \times \prod_{a=1}^{n_p} \Z_{m_a}$, while the matter content consists of $n$ fields $\Phi_i$ and $n_p$ fields  $P_a$, carrying $U(1)$ charges $+1$ and $-k_a$ respectively. The superpotential is
\begin{equation}
W = \sum_{a=1}^{n_p} P_a^{m_a} G_{m_ak_a}(\Phi)\, .
\end{equation}
where $G_{m_ak_a}(\Phi)$ is the most general degree $m_ak_a$ polynomial in $\Phi_i$.
The Lagrangian has a discrete symmetry
\begin{equation}
P_a \rightarrow e^{2\pi i /m_a} P_a\, , \ \forall a=1,\cdots,n_p\, ,
\end{equation}
which we declare to be a gauge symmetry. Here $m_a, k_a \geq 1$ and are
integral with the condition that if $m_a = 1$ then $k_a \geq 2$. 
Note that this class of GLSMs allows the ``$P$ fields'' to appear with
general exponents in the superpotential, in contrast to the standard
GLSM construction for complete intersections in toric varieties.
The R-charge of $\Phi_i$ is assumed to be $\rr_\phi>0$ and consequently $P_a$ has R-charge $\rr_a = 2/m_a - k_a \rr_\phi$. The anomaly and central charge conditions read
\begin{equation}
n = \sum_{a=1}^{n_p} k_a\, , \qquad
3 = n-1+n_p - 2\sum_{a=1}^{n_p} \frac{1}{m_a}\, . \label{eq:special-ac}
\end{equation}
The discrete gauge symmetry is important to ensure that gauge-invariant chiral operators carrying fractional left/right-moving R-charges, {\it e.g.}, $\cO = P_a \Phi_i^{k_a}$, are projected out. 

We apply the criteria in Table \ref{tab:gpc} and look for geometries in the $r \gg 0$ phase, {\it i.e.}, in the neighborhood of the singular point at $z =0$ (recall that $z= e^{-2\pi r+i\theta}$).
The $S^2$  partition function for this class of GLSMs is given by
\begin{align}
Z_{S^2} = \sum_{m\in \Z} \int_{\frac{\rr_\phi}{2}-i\infty}^{\frac{\rr_\phi}{2}-i\infty} \ \ \frac{d\tau}{2\pi i} \ z^{\tau - m/2} \ \zb^{\ \tau +m/2} \ \frac{\Gamma( -\tau - \frac{m}{2})^n}{\Gamma(1+\tau -\frac{m}{2})^n}\ \prod_{a=1}^{n_p} \frac{\Gamma(\frac{1}{m_a}+k_a( \tau + \frac{m}{2}))}{\Gamma(1-\frac{1}{m_a} + k_a( -\tau + \frac{m}{2}))}\, .
\end{align}
Near $z=0$, the partition function can be evaluated by closing the contour in the right half plane, thereby receiving contributions from the poles at $\tau = -m/2 + n$, with $n\geq 0$, $m\leq n$. This series of poles has order $n-\ell$, where $\ell$ denotes the number of $P_a$ with exponent $m_a=1$. (The $\Gamma$ functions associated to these fields produce zeros that reduce the pole order from $n$.) The geometric criteria imply that 
\begin{equation}
n-\ell = 4\, . \label{eq:special-MUM}
\end{equation}
There are only finitely many GLSMs in this class that satisfy conditions \eqref{eq:special-ac} and \eqref{eq:special-MUM}. The possibilities are listed in Table \ref{tab:soln-1}.

\begin{table}[t]
\centering
\begin{tabular}{|c|c|c|c|}
\hline
$n$ & $n_p$ & $P$-field charges ($k_a$) & Exponents in $W$ ($m_a$) \\
\hline
4   &   4   &  $(-1,-1,-1,-1)$ & $(2,2,2,2)$ \\
\hline
4   &   3   &  $(-2,-1,-1)$    & $(2,2,2)$ \\
5   &   4   &  $(-2,-1,-1,-1)$ & $(1,2,2,2)$ \\
\hline
4   &   2   &  $(-3,-1)$       & $(2,2)$ \\
5   &   3   &  $(-3,-1,-1)$    & $(1,2,2)$ \\
4   &   2   &  $(-2,-2)$       & $(2,2)$ \\
5   &   3   &  $(-2,-2,-1)$    & $(1,2,2)$ \\
6   &   4   &  $(-2,-2,-1,-1)$ & $(1,1,2,2)$ \\
\hline
4   &   1   &  $(-4)$          & $(2)$ \\
5   &   2   &  $(-1,-4)$       & $(2,1)$ \\ 
5   &   2   &  $(-2,-3)$       & $(2,1)$ \\
5   &   2   &  $(-2,-3)$       & $(1,2)$ \\
6   &   3   &  $(-2,-2,-2)$    & $(1,1,2)$ \\
6 	 &	  3   &  $(-3,-2,-1)$	 & $(1,1,2)$ \\
7	 &   4   &  $(-2,-2,-2,-1)$ & $(1,1,1,2)$ \\	
\hline
\hline
5   &   1   &  $(-5)$          & $(1)$ \\
6   &   2   &  $(-3,-3)$		 & $(1,1)$ \\
6   &   2   &  $(-2,-4)$		 & $(1,1)$ \\
7   &   3   &  $(-2,-2,-3)$	 & $(1,1,1)$ \\
8   &   4   &  $(-2,-2,-2,-2)$ & $(1,1,1,1)$ \\
\hline
\end{tabular}
\caption{Matter content of GLSMs satisfying the MUM criteria. The last five solutions are complete intersection Calabi--Yau varieties in $\P^{n-1}$.}
\label{tab:soln-1}
\end{table}

\subsubsection*{Is the $r>0$ phase geometric?}

This leads to the question of whether the singular point at $z=0$ is  a geometry. The partition function for a GLSM in Table \ref{tab:soln-1} is
\begin{align*}
Z_{S^2} = \sum_{m\in \Z} \int_{\frac{-\rr_\phi}{2}-i\infty}^{\frac{-\rr_\phi}{2}-i\infty} \ \frac{d\tau}{2\pi i} \ & e^{-4\, \pi\,  r \, \tau - i\, \theta\, m} \, \frac{\Gamma( -\tau - \frac{m}{2})^n}{\Gamma(1+ \tau -\frac{m}{2})^n}
 \prod_{\mu=1}^{\ell} \frac{\Gamma(1+k_\mu( \tau + \frac{m}{2}))}{\Gamma( k_\mu( -\tau + \frac{m}{2}))} 
 \prod_{\alpha = 1}^{n_p-\ell} \frac{\Gamma(\frac{1}{2} + k_\alpha( \tau + \frac{m}{2}))}{\Gamma(\frac{1}{2} + k_\alpha( -\tau + \frac{m}{2}))}\, .
\end{align*}
Note that we have split the product over $n_p$ fields depending on whether the corresponding $P_a$ have $m_a=1$ or $m_a=2$. 
Using $\Gamma$ function identities, the integrand can be transformed to 
\begin{align}
Z_{S^2} = \sum_{m\in \Z} \int_{\frac{-\rr_\phi}{2}-i\infty}^{\frac{-\rr_\phi}{2}-i\infty} \ \ \frac{d\tau}{2\pi i} \quad & e^{-4\, \pi\,  r' \, \tau - i\, \theta'\, m} \  \frac{\Gamma( -\tau - \frac{m}{2})^n}{\Gamma(1+ \tau -\frac{m}{2})^n}\  \ \prod_{\alpha = 1}^{n_p-\ell} \frac{\Gamma( -k_\alpha (\tau + \frac{ m}{2}))}{\Gamma(1+ k_\alpha(\tau -\frac{m}{2}))}
\nonumber \\
& \prod_{\mu=1}^{\ell} \frac{\Gamma(1 + k_\mu( \tau + \frac{m}{2}))}{\Gamma( k_\mu( -\tau + \frac{m}{2}))} \ \ 
 \prod_{\alpha = 1}^{n_p-\ell} \frac{\Gamma(1+2k_\alpha( \tau + \frac{m}{2}))}{\Gamma( 2k_\alpha( -\tau + \frac{m}{2}))}
 \, ,\label{eq:CI-2-cover}
\end{align}
where
\begin{equation}
r' = r + \frac{\log 2}{\pi} \sum_{\alpha=1}^{n_p-\ell} k_\alpha \, , \quad \theta' = \theta+\sum_{\alpha=1}^{n_p-\ell} k_\alpha \pi\, .
\end{equation}
We note that \eqref{eq:CI-2-cover} is precisely the partition function of a GLSM for a complete intersection of 
$n_p$ hypersurfaces, of degrees $k_{\mu=1}$, $\cdots$ , $k_{\mu=\ell}$ and $2k_{\alpha=1}$, $\cdots$, $2k_{\alpha=n_p-\ell}$, 
in the weighted projective space $Y=\P[1,1,\cdots,1, k_{\alpha=1}, \cdots, k_{\alpha=n_p-\ell}]$ of dimension $n+n_p-\ell-1=n_p+3$. The singular locus of the ambient space $Y$ is at least codimension $n$ and, consequently, does not intersect with the $n_p$ defining hypersurface equations. The complete intersection defines a smooth Calabi--Yau threefold geometry.

We have found that the partition function of each GLSM in the collection, as an
expansion around $z=0$, agrees with the partition function of a
geometric complete intersection, as an expansion around the large
volume point. The equality of the partition functions implies that the Gromov--Witten invariants in the $r>0$ phase must agree with those of the corresponding complete intersection Calabi--Yau threefold \cite{Jockers:2012dk}. This is evidence in favor of a geometric interpretation
of the $r>0$ phase of these GLSMs.  We will specialize to one solution
in Table \ref{tab:soln-1} and argue more convincingly for a geometric
interpretation of the $r>0$ phase. It is likely that this argument
generalizes to all the solutions in Table \ref{tab:soln-1}, but we
will restrict ourselves to this one example for the sake of
simplicity. 
\subsubsection*{The $P^2$-GLSM describes a geometry}
The GLSM we study, which we will refer to as the ``$P^2$-GLSM'',
has gauge group $U(1)\times \Z_2$ and matter consisting of superfields
$X_i$ with charge $+1$ and a field $P$ of charge $-4$; and under the
$\Z_2$ gauge symmetry, $P\rightarrow -P$. The superpotential takes the
form
\begin{equation}
W = P^2 G_8(X)\, ,
\end{equation}
where $G_8(x)$ is a generic homogeneous polynomial of degree 8.

We first analyze the phase structure of this GLSM. The classical vacuum moduli space is a solution of the equations
\begin{eqnarray}
\begin{array}{cc}
\mbox{D-term} & \mbox{F-terms} \\
\sum_{i=1}^4 |x_i|^2 - 4|p|^2 = r \quad & \quad p G_8(x) = 0\, , \quad p^2 \partial_iG_8 = 0.
\end{array}
\end{eqnarray}
When $r<0$ we have $p \neq 0$, which breaks the gauge group from $U(1)\times \Z_2$ down to a $\Z_8$, generated by the element $(e^{2\pi i/8}, -1) \subset U(1)\times \Z_2$. The $X_i$ are massless and interact through the effective superpotential $W = \langle P \rangle^2 G_8(X)$. This phase is a Landau-Ginzburg orbifold.

When $r>0$, we must have $x\neq 0$. Since $G_8(x)$ is generic we must have $p=0$. The classical vacuum moduli space is a $\P^3$ spanned by the $x_i$ with an unbroken $\Z_2$ gauge symmetry. However, this classical analysis is insufficient, as was pointed out in \cite{Aspinwall:1994cf, Caldararu:2007tc,Hellerman:2006zs} in a closely related example. Thinking semi-classically, although we have $\langle p \rangle=0$, the field becomes massless and fluctuates along the octic locus $(G_8(x)=0) \subset \P^3$. Away from this locus, $p$ can be integrated out, leaving an unbroken $\Z_2$ gauge symmetry. A $\Z_2$ gauge theory has two bosonic vacua, and thus the moduli space of vacua --- away from the octic locus --- consists of two copies of $\P^3$. This phase is referred to as a \emph{true hybrid} phase in the literature since the $\P^3$ is fixed by the R-symmetry \cite{Aspinwall:2009qy}. 
\begin{center}
\begin{tikzpicture}
\draw[thick] (-2,0)--(0,0) cos (2,-1) sin (4,-2)--(6,-2);
\draw[thick] (6,0)--(4,0) cos (2,-1) sin (0,-2)--(-2,-2);
\fill[color=gray] (2,-1) circle (.9cm);
\node at (2,-1) {$G_8(x)=0$} [];
\node at (5,-1.7) {$\P^3$} [];
\node at (5,-0.3) {$\P^3$} [];
\end{tikzpicture}
\end{center}
This picture suggests that this phase may be related to a Calabi--Yau threefold that is obtained by taking a double-cover of $\P^3$ branched over an octic surface. We have shown above that the $S^2$ partition function in this phase is equal to the partition function of a degree 8 hypersurface in $\P^{1,1,1,1,4}$ with homogeneous coordinates $[x_1,x_2,x_3,x_4,y]$. A \emph{special} family of hypersurfaces of the form
\begin{equation}
y^2 = G_8(x)\, ,
\end{equation}
demonstrates that this Calabi--Yau threefold, on a sublocus of its complex structure moduli space, can indeed be viewed as a double-cover of $\P^3$ branched over an octic surface. The weighted projective space $\P^{1,1,1,1,4}$ has singularities at points, which are missed by the generic hypersurface. For generic $G_8(x)$ the Calabi--Yau threefold is smooth.

We will now show that the $r>0$ phase is geometric. To do so we consider a new GLSM, which we will refer to as GLSM2, with gauge group $U(1)\times U(1)$ and matter content given below.
\begin{center}
\begin{tabular}{|c|cccc|}
\hline
Field: 	& $X_i$ & $Y$ & $Q$ & $P$ \\
\hline
$U(1)_1$ 	& $+1$  &  $4$ & $0$  & $-8$ \\
\hline
$U(1)_2$ 	& $0$   &  $1$ & $1$  & $-2$ \\
\hline
\end{tabular}
\end{center}
GLSM2 has a superpotential
\begin{equation}
W = P(Y^2 - Q^2G_8(X))\, .
\end{equation}
The classical vacuum moduli space is a solution to the equations
\begin{equation}
\begin{array}{rl}
\mbox{D-terms\qquad \qquad } & \mbox{\qquad \qquad \qquad F-terms} \\
\sum_{i=1}^4 |x_i|^2 + 4|y|^2 - 8|p|^2 = r_1\, , & \quad y^2 - q^2 G_8(x) = 0\, , \quad py=0 \, ,\\
|y|^2 + |q|^2 - 2|p|^2 = r_2\, ,     &   \quad  pqG_8(x)=0\, , \quad pq^2 \partial_i G_8 = 0\, .
\end{array} \label{eq:DF-GLSM2}
\end{equation}
The low-energy effective theory is a function of $(r_1,r_2)$ and is depicted in Figure \ref{fig:phases-GLSM2}. 
The phase boundaries in figure \ref{fig:phases-GLSM2} are
the one-dimensional rays generated by the charge vectors $\bQ_i=(1,0)$, $\bQ_Q=(0,1)$
and $\bQ_P=(-8,-2)$, agreeing  with the secondary fan discussion of section
\ref{sec:abelian phases}.\footnote{The secondary fan also contains the ray $\bQ_{Y}=(4,1)$, which is not a real phase boundary as far as the IR physics is concerned due to the F-terms. From the partition function perspective, the associated poles in $Z_Y$ are cancelled by zeroes of $Z_P$.}

We describe the phases as:
\begin{itemize}
\item \underline{$r_1>0,r_2>0$}: The equations \eqref{eq:DF-GLSM2} imply that  $q \neq 0$ and $x\neq 0$. Consequently, we have $p=0$ for generic $G_8(x)$. Since $q\neq 0$ we can use the $\C^*$ action associated with $U(1)_2$ to set $q=1$. The vacuum moduli space is given by the Calabi--Yau hypersurface $X$: $(y^2 - G_8(x)=0)$ in $\P^{1,1,1,1,4}$, which is a double-cover of $\P^3$ branched over the octic $G_8(x)=0$. 

\item \underline{$r_2<0, r_1-4r_2>0$}: We have $x\neq 0$ and $p\neq 0$, which implies that $y=0$ and $q=0$. Using the D-term we can set $\langle p \rangle = \sqrt{|r_2|/2}$. The field $Y$ can be integrated out since it has a mass proportional to $\sqrt{|r_2|}$. The low-energy effective theory consists of the fields $Q$ and $X_i$ interacting via a superpotential
\begin{equation}
W_{eff} = \sqrt{|r_2|/2} \ Q^2 G_8(X)\, .
\end{equation}
The expectation value of $P$ breaks the gauge group from $U(1)\times U(1)$ to $U(1)\times \Z_2$, where the $Z_2$ acts as $Q \rightarrow -Q$. This is precisely the $r>0$ phase of the $P^2$-GLSM! We will denote this phase by $Y$, which can be imagined as two copies of $\P^3$ ``glued together'' by an interacting Landau-Ginzburg theory.

\item \underline{$r_1<0, r_1-4r_2 <0$}: We have $q\neq 0$ and $p\neq 0$. The gauge group is broken to $\Z_8$ and we have the Landau-Ginzburg orbifold phase that we encountered earlier as the $r<0$ phase of the $P^2$-GLSM.
\end{itemize}
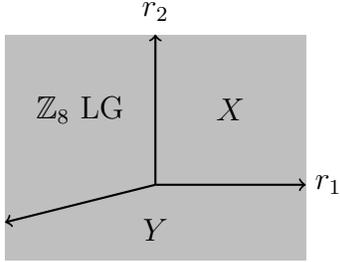
\begin{figure}
\centering
\begin{tikzpicture}
\fill[color=lightgray] (-2,-1) rectangle (2,2);
\draw[thick, ->] (0,0)--(2,0);
\node at (2.3,0) {$r_1$} [];
\draw[thick,->] (0,0)--(0,2);
\node at (0,2.3) {$r_2$} [];
\draw[thick,->] (0,0)--(-2,-0.5);
\node at (1,1) {$X$} [];
\node at (0,-0.6) {$Y$} [];
\node at (-1,1) {$\Z_8$ LG} [];
\end{tikzpicture}
\caption{Phases of GLSM2.}
\label{fig:phases-GLSM2}
\end{figure}

GLSM2 in the $r_1>0,r_2>0$ phase describes a hypersurface in a toric
variety, which is a resolution of the singularities of the weighted
projective space $\P_{1,1,1,1,4}$. Since the generic hypersurface
misses the singular points in $\P_{1,1,1,1,4}$ it is insensitive
to the details of the resolution and therefore the IR SCFT associated
to this phase of GLSM2 must depend on only one combination
of FI parameters. The two-sphere partition function must reflect this fact,
 and upon computing it we find that the relevant combination is
\begin{equation}
w = \frac{2^8 z_1}{(1-4z_2)^4}\, .
\end{equation}
We emphasize that the IR physics is the same along the loci $w=$ constant.
Since one such locus connects the boundary points associated to phases
$X$ (the geometric phase) and $Y$ ($P^2$-GLSM hybrid phase), these phases
have identical IR physics and therefore the hybrid theory describes a geometry.

We believe that a similar analysis will hold for  the GLSMs listed in Table \ref{tab:soln-1},
and therefore they should describe Calabi--Yau threefold geometries in the $r>0$ phase.

\section{GKZ systems for GLSMs}
\label{sec:GKZ}
\def\cT{\mathcal{T}}
\def\cL{\mathcal{L}}
\def\xb{\bar{x}}
\def\tauk{\bm{\tau}_{\bm{k}}}
\def\bme{\bm{\epsilon}}

In this section we prove that the two-sphere partition function of any abelian GLSM satisfies a system of differential equations (\ref{eq:GKZ1}, \ref{eq:GKZ2}) that are a generalization of the $A$-hypergeometric system defined by Gelfand, Kapranov and Zelevinski (GKZ) \cite{GKZ1, GKZ2}. When the abelian GLSM corresponds to a complete intersection in a toric variety this set of differential equations is precisely the $A$-hypergeometric system. We know from special geometry that the partition function (or $e^{-K}$) is a symplectic combination of the holomorphic periods and their complex conjugates. The periods themselves must therefore be a solution to this system of differential equations.

Consider a general abelian GLSM with $G=U(1)^s \times \Gamma$ and $n$ chiral fields $\Phi_{i=1, \cdots, n}$ with charges $\bm{Q}_i$ and R-charge $\rr_i$ . $\Gamma$ is a discrete abelian gauge group factor that acts on the fields $\Phi_i$ through multiplication by a phase $e^{i 2\pi \varphi_i(\gamma)}$, where $\gamma \in \Gamma$ is a group element and $\varphi_i(\gamma) \in \mathbb{Q}$. The set $\cT$ of $G$-invariant Laurent monomials
\begin{equation}
\cT = \big\{ \prod_{i=1}^n\phi_i^{t_i} \ | \ (t_i) \in \Z^n\, , \ \sum_i t_i \bm{Q}_i = 0 \, , \ \sum_{i=1}^n \varphi_i(\gamma) t_i \in  \Z\,  \big\}\, ,
\end{equation}
is parameterized by a lattice that we will denote by $M \cong \Z^{n-s}$, following \cite{summing}. This parameterization can be defined in terms of a set of vectors $A := \{ v_1, \cdots, v_n\}$  in the dual lattice $N := \mathrm{Hom}(M,\Z)$ as 
\begin{equation}
\cT = \big\{ \prod_{i=1}^n\phi_i^{v_i(m)} \ | \ m \in M\,  \big\}\, .
\end{equation}
Let $L := \{ \ (l_i) \in \Z^n\ | \ \sum_{i=1}^n l_i v_i = 0 \}$, denote the lattice of relations amongst the $n$ vectors in $A$. Note that this lattice $L$ is generated over the integers by the vectors $\{Q_{ia}\}_{a=1, 2, \cdots, s}$. For any $\beta \in N \otimes \C$, we define a set of differential equations --- which we refer to as the $A$-system --- for a function $\Psi(x_1, \cdots, x_n)$:
\begin{eqnarray}
\sum_{i=1}^n v_i(m) \ x_i \frac{\partial}{\partial x_i} \Psi = \beta(m) \Psi\, , \quad \forall \ m \in M\, , \label{eq:GKZ1} \\
\prod_{l_i > 0} \left( \frac{\partial}{\partial x_i} \right)^{l_i} \Psi = \prod_{l_i <0} \left( \frac{\partial}{\partial x_i} \right)^{|l_i|} \Psi \, , \quad \forall \ l \in L \, . \label{eq:GKZ2}
\end{eqnarray}
The
$A$-system is a slight generalization of the GKZ system since we do not
impose the condition (see \cite{GKZ1, GKZ2}) that $v_i$ generate the lattice $N$, or that
there should exist an element $\mathrm{deg} \in M$ such that
$\deg(v_i)=1$ for $i=1, \cdots, n$.

Our moduli space $\cM$ is isomorphic to $(\mathbb{C}^*)^s$ and
has a natural compactification $\overline{\cM}$
specified by the secondary fan of the GLSM.  This compactification uses
the traditional description of a toric variety in terms of a fan in
a real vector space (see, for example, \cite{catp} for an exposition of
this approach), and the boundary points produced in this way include a
point of maximal codimension associated to each GLSM phase around which
we can expand physical quantities such as $Z_{S^2}$.

A more familiar description of $\overline{\cM}$ is as a holomorphic quotient
\cite{cox}, and this construction of it will also allow us to describe
the GKZ system.
We introduce an auxiliary space $\cV \cong \C^n$ with coordinates $x_{i=1, \cdots, n}$ along with a $(\C^*)^{n-s}$ action: $x_i \sim  \lambda_i^{v_{i\alpha}}\ x_i$, where $v_i\in A$. The variables $x_i$ are the homogeneous coordinates of the toric variety\footnote{%
Here $\cZ$ is a closed subset
of $\cV$ determined by the fan, which must be removed before the quotient
is taken.  See \cite{catp} for an explanation of how this works in detail.}
 $\overline{\cM}= (\cV - \cZ)/(\C^*)^{n-s}$ whose fan is the secondary fan corresponding to the abelian GLSM. 
The FI parameters are specified in terms of the $x_i$ as
\begin{equation}
z_a = \prod_{i=1}^n \ x_i^{Q_{ia}}\, .
\end{equation}
The partition function, which is a real-valued function on the space of FI parameters $(\C^*)^s$, can be extended to a function $\Psi(x,\xb)$ on the space $\cV$ as 
\begin{eqnarray}
\Psi(x, \xb) & = & \left( \prod_{i=1}^n (x_i\xb_i)^{-\frac{\rr_i}{2}} \right) Z_{S^2}( \prod_{i=1}^n \ x_i^{Q_{ia}},  \prod_{i=1}^n \ \xb_i^{Q_{ia}}) \nonumber \\
& = & \sum_{\bm{m} \in \Z^s} \ \int \prod_{a=1}^s \frac{d\tau_a}{2\pi i} \ \prod_{i=1}^n x_i^{-\frac{\rr_i}{2} + \bm{Q}_i\cdot (\tau - \frac{m}{2})}\ \xb_i^{-\frac{\rr_i}{2} + \bm{Q}_i\cdot (\tau + \frac{m}{2})} 
 \frac{\Gamma(\frac{\rr_i}{2} - \bm{Q}_i \cdot(\bm{\tau} + \frac{\bm{m}}{2}) )}{\Gamma(1-\frac{\rr_i}{2} + \bm{Q}_i\cdot (\bm{\tau} - \frac{\bm{m}}{2}) )} \, . \label{eq:psi-def}
\end{eqnarray}
Note that the function $\Psi$ is singular on the compactification $\overline{\cV}$ 
as expected, since the metric on moduli space is singular at the boundaries. 
Consider a particular phase  of the GLSM defined by  $\sum_{b=1}^s \ p_{ab} r_b > 0$, which corresponds to a singular point at $\xi_a=0$, $a=1, \cdots, s$, where $\xi_a:= \prod_{b=1}^s\ z_b^{p_{ab}}$. The partition function in this phase is a function of the variables $(\xi, \xib)$. The prefactor in \eqref{eq:psi-def} can be expressed as $|\prod_{a=1}^s \xi_a^{\delta_a}|^2$, where $\delta_a \in \mathbb{Q}$, using the $(\C^*)^{n-s}$ action on $\cV$. Therefore,
$\Psi$ agrees with $Z_{S^2}$ up to a K\"ahler transformation and hence contains the same physics as the partition function. 

Although $x_i$ and $\xb_i$ are related by complex conjugation we regard them as independent variables. $\Psi(x,\xb)$ will therefore be viewed as a holomorphic function in the variables $x_i$ and $\xb_i$. 

\vspace{0.1in}
\noindent
{\bf Theorem}: {\it $\Psi(x,\xb)$ satisfies the $A$-system of differential equations \eqref{eq:GKZ1}, \eqref{eq:GKZ2}, with 
$\beta = -\frac{1}{2} \sum_{i=1}^n \rr_i v_i$. }

\vspace{0.1in}

For a proof of this assertion see Appendix \ref{sec:GKZ ap}.
The $A$-system, which $\Psi(x,\ov x)$ satisfies, is a generalization of
the $A$-hypergeometric system defined by GKZ. For some GLSMs the $A$-system
is precisely a GKZ system; this includes any abelian GLSM with a geometric
phase describing a Calabi--Yau threefold which is a complete intersection
in a toric variety (CICY). However, we have shown that one can associate
an $A$-system to \emph{any} GLSM with an abelian gauge group ---
a broader class of GLSMs than those that describe CICYs.  For SCFTs that have integral left and right moving R-charges, $\beta$ is an integral vector since every element of the lattice $\cT$ must have even vector R-charge.

As a simple example of an $A$-system which is not of GKZ type, consider the $P^2$-GLSM from the previous section. The gauge group is $U(1)\times \Z_2$ with chiral fields $X_{1, 2, 3, 4}$ with charge $+1$ and a field $P$ of charge $-4$. The $\Z_2$ gauge symmetry acts as $P \rightarrow -P$. The superpotential is given by $W = P^2 G_8(X)$, and we can assign the R-charges $\rr_x$ to $X_{1,2,3,4}$ and R-charge $1-4\rr_x$ to $P$. The lattice of $G$-invariant Laurent monomials can be parameterized in terms of a lattice $M \cong \Z^4$ as
\begin{equation}
\cT = \{ X_1^{2m_1+m_2} X_2^{2m_1+m_3} X_3^{2m_1+m_4} X_4^{2m_1-m_1-m_2-m_3} P^{2m_1}\ | \ \ m \in M = \Z^4 \, \}\, ,
\end{equation}
which determines the set $A = \{ v_1, v_2, v_3, v_4, v_p\} \subset N \cong \Z^4$ as follows:
\begin{equation}
A = \{ (2,1,0,0)\, , \ (2,0,1,0)\, , \ (2,0,0,1)\, , \ (2,-1,-1,-1)\, , \ (2,0,0,0)\, \}\, .
\end{equation}
Note that the vectors in $A$ do not generate the lattice $N$, and lie on a hyperplane that has normal distance 2 from the origin $(0,0,0,0)$. The lattice of relations is given by
\begin{equation}
L = \{ (l , l , l,l,-4l)\, | \ l \in \Z \}\, ,
\end{equation}
and the vector $\beta = (-1,0,0,0)$.

In section \ref{sec:P2} we argued that this GLSM actually describes a Calabi--Yau threefold geometry --- a special degree 8 hypersurface in $\P^{1,1,1,1,4}$ --- in one of its phases. Mirror symmetry provides a description of the quantum K\"ahler moduli space in this case. The Picard-Fuchs operator is given by \cite{Morrison:1991cd}
\begin{equation}
\cL_{PF} = (w-256) \left( w \frac{\partial}{\partial w} \right)^4  + 2w\left( w \frac{\partial}{\partial w} \right)^3+\frac{43w}{32} \left( w \frac{\partial}{\partial w} \right)^2 + \frac{11w}{32} \left( w \frac{\partial}{\partial w} \right)+\frac{105w}{4096} \label{eq:pf-double-cover}
\end{equation}
 in terms of an algebraic coordinate $w$ on the moduli space.
The $A$-system defined above reduces to a single order 4 linear differential operator in the coordinate $z = \frac{x_1x_2x_3x_4}{{x_p}^4}$:
\begin{equation}
z \prod_{n=0}^3 \left( -\frac{2n+1}{2} +  z \frac{\partial}{\partial z} \right) - \left( z\frac{\partial}{\partial z} \right)^4\, .
\end{equation}
It is easily checked that this differential operator agrees with the Picard-Fuchs operator \eqref{eq:pf-double-cover} derived using mirror symmetry, if the coordinates $z$ and $w$ are related as $w = \frac{1}{256 z^4}$.

As in our discussion of phases, the associated Cartan theory provides some information about the periods of a general non-abelian GLSM. The associated Cartan theory, being an abelian GLSM, can be associated with an $A$-system. This system of differential equations can be solved in a given GLSM phase yielding a complete set of solutions. On restriction to the Weyl-invariant subspace of the FI parameter space, we expect the periods of the non-abelian theory to be contained in the space of solutions. We postpone a more detailed study of the non-abelian GLSM along these lines to future work.


\section{Conclusions}
In this paper we have utilized the two-sphere partition function
$Z_{S^2}$ to study three aspects of GLSM physics: the appearance
of classical phase transitions upon movement in the space of FI parameters,
criteria necessary for the existence of a geometric phase, and
a system of differential equations on the moduli space of the IR SCFT.

We saw in section \ref{sec:phases} that the analytic structure of
the partition function integrand recovers the description of abelian
GLSM phases in terms of the secondary fan \cite{MR1020882,BFS,Aspinwall:1993nu}.
Passing through codimension one walls in FI space moves the theory
to a different maximal cone of the secondary fan; this requires a different choice of contours in the partition
function integral, and the 
phase transition is detected through a change in the structure of
residues. Some non-abelian phase transitions can be detected by
studying the secondary fan of the associated Cartan theory, as demonstrated
in the Gulliksen-Neg\r{a}rd GLSM.

In section \ref{sec:geometries} we developed a set of \emph{geometric
  phase criteria}, which a GLSM phase must satisfy in order to admit a
description in terms of a Calabi--Yau non-linear sigma model. 
These criteria, which include conditions on the partition function, can be easily checked in a given GLSM phase.
We formulated another potential criterion and demonstrated that its
violation suggests the existence of singularities in the geometry,
using hypersurfaces in weighted projective space as an example.  We
also constructed a GLSM which has a phase satisfying all the criteria, but
is actually singular due to the existence of non-compact directions in
the moduli space of vacua. This shows that the geometric phase criteria are
necessary but not sufficient for the existence of a geometric phase.

In section \ref{sec:illustrative examples} we applied the geometric
phase criteria to identify a set of novel geometric phases in GLSMs. We 
studied a collection of GLSMs with gauge group $U(1)$ that satisfy the
criteria. Focusing on one member of the collection, we showed
that the relevant phase looks like a hybrid: two copies of $\bP^3$ glued together by a Landau--Ginzburg model. 
We then realized the same phase in an auxiliary two-parameter GLSM and showed that the hybrid phase is actually geometric, using the following physical argument ---  the IR physics in the auxiliary GLSM is the same along codimension-one loci in FI space, and one such locus connects the hybrid phase of interest to a geometric phase. The Calabi--Yau threefold realized by the hybrid phase is a double-cover of $\P^3$ branched over an octic surface.

In section \ref{sec:GKZ} we demonstrated that a system of linear differential equations, which we call the $A$-system, 
can be associated to \emph{any} abelian GLSM, and $Z_{S^2}$ is a solution of
this system. 
For
abelian GLSMs which describe complete intersections in a toric
variety, this system reduces to the $A$-hypergeometric system of Gel'fand, Kapranov and Zelevinski.  We leave an investigation along these lines for non-abelian theories to future work.

Another interesting direction for future work is to apply the method
of section \ref{sec:illustrative examples} to large collections of GLSMs
specified by the gauge group, matter representations and
R-charges. The geometric phase criteria would then give us a set of
candidate Calabi--Yau threefold geometries. Upon closer study, some of
these GLSMs could lead to novel constructions of Calabi--Yau
threefolds.

{\noindent \bf Acknowledgments.} We thank F.~Benini, M.~Cveti\v{c}, R.~Donagi, J.~Gomis, R.~Gopakumar, A.~Grassi, K.~Hori, S.~Hosono, N.~Iqbal, H.~Jockers, D.~Klevers, J.~M.~Lapan, B.~Le~Floch, S.~Lee, L.~McAllister, R.~Plesser, M.~Romo, A.~Sen, and S.~Trivedi for useful discussions and 
S.~K.~Harris and J.~M.~Lapan for comments on a draft. We are especially grateful to J.~M.~Lapan for extensive discussions on the $P^2-$GLSM and related issues. V.~K.~would like to thank the Tata Institute for Fundamental Research, the Harish-Chandra Research Institute, for hospitality while this research was being carried out, and the participants of the International Strings Meeting 2012, Puri, India for interesting discussions. V.~K.~and D.~R.~M.~are grateful to the organizers and participants of the workshop ``Seminar weeks on Calabi-Yau manifolds, mirror symmetry, and derived
categories'' at the University of Tokyo for useful discussions, and D.~R.~M. is equally grateful to the organizers and participants of the meeting ``New Mathematical Structures in Supersymmetric Gauge Theory?'' at the Perimeter Institute for Theoretical Physics.  
The work of J.~H.~and V.~K.~was supported in part by the National Science Foundation under Grant No. PHY11-25915, and
the work of D.~R.~M. was supported in part by National Science Foundation Grant DMS-1007414.

\appendix

\section{GKZ Proof}
\label{sec:GKZ ap}
The property $\sum_{i=1}^n \bm{Q}_i v_i = 0$ is sufficient to demonstrate that \eqref{eq:GKZ1} is satisfied. 

We now show that $\Psi(x,\ov x)$ as defined in (\ref{eq:psi-def}) is a solution to
the set of differential operators \eqref{eq:GKZ2}.  
In any given phase of the
GLSM the $s$-dimensional contour integral can be expressed as a sum of
residues at points that lie inside the contour of integration. The
integral naturally takes the form of an infinite power series, and we
will verify equation \eqref{eq:GKZ2} term wise by comparing
coefficients.

Consider the contribution to the integral \eqref{eq:psi-def} from a single pole located at the intersection of $s$ pole hyperplanes $H^{(k_\alpha)}_\alpha$, for $\alpha\in I$, where $I \subset \{1, \cdots, n\}$ is an $s$-element subset, with $\{ \bm{Q}_\alpha\}_{\alpha \in I}$ linearly independent. For each $\bm{m} \in \Z^s$, we study the contribution from the pole located at $\bm{\tau} = -\bm{m}/2+ \tauk$, which is the unique solution of the equations
\begin{equation}
\frac{\rr_\alpha}{2} - \bm{Q}_\alpha \cdot (\bm{\tau} + \frac{\bm{m}}{2}) = -k_\alpha, \quad \mbox{for } \alpha \in I\, .
\end{equation}
The residue contribution, for a fixed $\bm{m} \in \Z^s$, is
\begin{equation}
\Psi_{\bm{k}, \bm{m}} = \Res_{\bm{\epsilon}=0}\ \  \prod_{i=1}^n x_i^{-\frac{\rr_i}{2} + \bm{Q}_i\cdot (\tauk+\bm{\epsilon} - \bm{m})} \xb_i^{-\frac{\rr_i}{2} + \bm{Q}_i\cdot (\tauk+\bm{\epsilon})} \frac{\Gamma(\frac{\rr_i}{2} - \bm{Q}_i\cdot(\tauk+\bm{\epsilon}))}{\Gamma(1-\frac{\rr_i}{2}+\bm{Q}_i\cdot(\tauk+\bm{\epsilon}-\bm{m}))}\, .
\end{equation}
Since the lattice $L$ is generated by the vectors $Q_{i,a=1}$, $Q_{i, a=2}$, $\cdots$, $Q_{i,a=s}$,  we can express $l_i = Q_{ia}p_a$, where $p_a \in \Z$. 
Using standard $\Gamma$-function identities we have
\begin{eqnarray}
\prod_{ l_i>0} \left( \frac{\partial}{\partial x_i} \right)^{l_i} \Psi_{\bm{k},\bm{m}} &  =&   \Res_{\bm{\epsilon}=0}\ \  \prod_{i=1}^n x_i^{-\frac{\rr_i}{2} + \bm{Q}_i\cdot (\tauk+\bm{\epsilon} - \bm{m})} \xb_i^{-\frac{\rr_i}{2} + \bm{Q}_i\cdot (\tauk+\bm{\epsilon})} 
\frac{\Gamma(\frac{\rr_i}{2} - \bm{Q}_i\cdot(\tauk+\bm{\epsilon}))}{\Gamma(1-\frac{\rr_i}{2}+\bm{Q}_i\cdot(\tauk+\bm{\epsilon}-\bm{m}))} \nonumber \\
& & \prod_{l_j>0} \frac{1}{x_j^{l_j}} \frac{\Gamma(1 - \frac{\rr_j}{2}+\bm{Q}_j\cdot(\tauk+\bme - \bm{m}))}{\Gamma(1- \frac{\rr_j}{2}+\bm{Q}_j\cdot(\tauk+\bme - \bm{m})-l_j)}\, . \nonumber \\
& = & \Res_{\bme=0} \prod_{i=1}^n  x_i^{-\frac{\rr_i}{2} + \bm{Q}_i\cdot (\tauk+\bm{\epsilon} - \bm{m})-l_i} \xb_i^{-\frac{\rr_i}{2} + \bm{Q}_i\cdot (\tauk+\bm{\epsilon})}
\frac{\Gamma(\frac{\rr_i}{2} - \bm{Q}_i\cdot(\tauk+\bm{\epsilon}))}{\Gamma(1-\frac{\rr_i}{2}+\bm{Q}_i\cdot(\tauk+\bm{\epsilon}-\bm{m})-l_i)} \nonumber \\
& & \prod_{l_j < 0}  \frac{1}{x_j^{|l_j|}} \frac{\Gamma(1 - \frac{\rr_j}{2}+\bm{Q}_j\cdot(\tauk+\bme - \bm{m})-l_j)}{\Gamma(1- \frac{\rr_j}{2}+\bm{Q}_j\cdot(\tauk+\bme - \bm{m}))}\, . \nonumber \\
& =&  \prod_{l_i <0}  \left( \frac{\partial}{\partial x_i} \right)^{|l_i|} \Psi_{\bm{k},\bm{m}+ \bm{p}} \, .
\end{eqnarray}
Now, the contribution to $\Psi$ from the pole located at $\bm{\tau} = -\bm{m}/2+ \tauk$ is obtained by summing over all $\bm{m} \in \Z^s$, and then summing over all contributing poles. Through a relabeling $\bm{m} \rightarrow \bm{m} - \bm{p}$, it is easily seen that
\begin{equation}
\prod_{ l_i>0} \left( \frac{\partial}{\partial x_i} \right)^{l_i} \sum_{\mathrm{poles}} \ \sum_{\bm{m} \in \Z^s} \Psi_{\bm{k},\bm{m}} = \prod_{ l_i< 0} \left( \frac{\partial}{\partial x_i} \right)^{|l_i|}  \sum_{\mathrm{poles}} \  \sum_{\bm{m} \in \Z^s} \Psi_{\bm{k},\bm{m}}\, .
\end{equation}
Therefore, we see that equation \eqref{eq:GKZ2} is true in any phase of the GLSM. $\hspace{1cm}\blacksquare$

\section{Evaluation of the Partition Function}
\label{sec:comments and grid}

The discussion of section \ref{sec:geometries} discussed contributions
to $Z_{S^2}$ from point sets $P_I$. At a given point $\br$ in the
space of FI-parameters, there may be many such sets, and an important
practical consideration is how to set up sums in the partition function
which ensure that all contributing points are identified.
Though there are likely better methods, a simple method is to set up a ``grid.''

To illustrate the idea, consider a simple example.
Take the $\bP^4_{1,1,2,2,2}$ GLSM, which has two fields $\phi_i$ with $Q_i=1$
and $\rr_i=0$, three fields $\chi_i$ with $Q_i=2$ and $\rr_i=0$, and one field $P$
with $Q_i = -8$ and $\rr_i=2$. The one-loop determinants are given by
\begin{equation}
Z_\phi = \frac{\Gamma(-\tau - \frac{m}{2})}{\Gamma(1+\tau-\frac{m}{2})}\qquad \qquad 
Z_\chi = \frac{\Gamma(-2\tau - m)}{\Gamma(1+2\tau-m)} \qquad \qquad 
Z_P = \frac{\Gamma(1+8\tau+4m)}{\Gamma(-8\tau +4m)}.
\end{equation}
The $\Gamma$-functions in the numerators define infinite
sets of polar hyperplanes where the arguments are negative integers.
The points $p$ in $\tau$-space which are in the intersection of many
such hyperplanes have $\tau_a \in \bQ$ for all $a$, and for simplicity we set up a summation
which contains all contributing points $p$ in all phases. However, this requires
a careful choice of summation variable: if one were to sum over
$\tau = k - \frac{m}{2}$ with $k \in \bZ$, all $Z_\phi$ poles but only
every second $Z_\chi$ pole and every eighth $Z_P$ pole would be in the
sum. Instead, if one sums over $\tau = \frac{k}{n} - \frac{m}{2}$ with
$k \in \bZ$ and $n=8$ every pole would be in the sum. More
generically, a similar expansion holds for any $\mathbb{W}\bP^4$ GLSM
with $n = lcm(F)$, $F=\{Q_i\}$.

It is easy to see how the one-dimensional grid just described generalizes
to a grid which captures all contributing points $p$ for any GLSM\footnote{Both abelian
and non-abelian since the poles of $Z_{S^2}^{NA}$ are a subset of $Z_{S^2}^C$ poles.}.
Defining the sets of charges $F_a = \{Q_{ia}\}$ and $n_a = lcm(F_a)$, sum over
$\tau_a = \frac{k_a}{n_a\, K} - \frac{m_a}{2}$ with $k_a \in \bZ$, where $K$ depends
on the R-charges. For example, for any GLSM with integer R-charges $\rr_i$, $K=2$ suffices.
Having such a sum over poles, one integrates in neighborhoods of the poles defined
by $\tau_a + \ep_a = \frac{k_a}{n_a K} - \frac{m_a}{2} + \ep_a$ and the integrals take the
form $\oint \dots \oint \frac{d\ep_1}{2\pi i} \cdots \frac{d\ep_s}{2\pi i}$. Rewriting
$\Zc$ in terms of these new variables, and also replacing $m_a$ via defining a new variable
$l_a := \frac{k_a}{n_a K} - m_a$, we obtain
\begin{equation}
  \Zc = e^{-4\pi \br \cdot \bt - i \bth \cdot \bm{m}} = \prod_a (z_a \ov z_a)^{\ep_a} \,\, z_a^{l_a}\,\, \ov z_a^{\frac{k_a}{n_a K}}.
\end{equation}
Note that unless $k_a = 0 \,\,\text{mod} (n_a\, K) \,\,\,\,\, \forall a$,
some number of the $z_a$ variables will be fractional. If there are non-trivial poles
for such $k_a$, the fractional powers appear in the partition function $Z_{S^2}$, and 
in some examples we will see that fractional powers are related to singular geometries \ref{sec:resolution}.
Note also that two distinct sets of $k_a$'s which are equivalent $mod(n_a K)$ for all $a$ define points in $\tau_a$
space with the same structure of polar divisors. This defines a ``modular box''.
Finally, since the summation variables $k_a$ have been set up in a phase independent
manner, the choice of a phase generically requires that the $k_a$ satisfy certain
inequalities governed by the choice of contours. This is simple to address on
an example by example basis.

\bibliographystyle{utphys}
\bibliography{refs}

\end{document}